\pdfoutput=1
\documentclass[fleqn,usenatbib]{mnras} 

\usepackage{newtxtext,newtxmath}

\usepackage[T1]{fontenc}
\usepackage{ae,aecompl}

\usepackage{graphicx}	
\usepackage{amsmath}	
\usepackage{amssymb}	
\usepackage{booktabs,dcolumn} 
\usepackage{threeparttable}
\usepackage{enumitem}
\setlist[description]{leftmargin=\parindent,labelindent=\parindent}

\usepackage{soul} 

\newcommand{\quotes}[1]{``#1''}
\newcommand*\rfrac[2]{{}^{#1}\!/_{#2}} 

\title[Witnessing the early stages of SSCs formation]{Super Hot Cores in NGC 253: Witnessing the formation and early evolution of Super Star Clusters}

\author[F. Rico-Villas  et al.]{
F. Rico-Villas,$^{1}$\thanks{E-mail: fernando.rico@cab.inta-csic.es}
J. Mart\'in-Pintado$^{1}$,
E. Gonz\'alez-Alfonso$^{2}$,
S. Mart\'in$^{3, 4}$
\newauthor
and V. M. Rivilla$^{5}$
\\
$^{1}$Centro de Astrobiolog\'ia (CSIC-INTA). Ctra de Ajalvir, km. 4, Torrej\'on de Ardoz, 28850, Madrid, Spain\\
$^{2}$Universidad de Alcal\'a, Departamento de F\'isica y Matem\'aticas, Campus Universitario, Alcal\'a de Henares, 28871, Madrid, Spain\\
$^{3}$European Southern Observatory, Alonso de C\'ordova, 3107, Vitacura, Santiago 763-0355, Chile\\
$^{4}$Joint ALMA Observatory, Alonso de C\'ordova, 3107, Vitacura, Santiago 763-0355, Chile\\
$^{5}$INAF-Osservatorio Astrofisico di Arcetri, Largo Enrico Fermi 5, 50125, Florence, Italy
}
\date{Accepted 2019 November 25. Received 2019 November 17; in original form 2019 September 25}
\pubyear{2019}

\begin{document}
\label{firstpage}
\pagerange{\pageref{firstpage}--\pageref{lastpage}}
\maketitle

\begin{abstract}
Using $0.2^{\prime \prime}$ ($\sim3$\,pc) ALMA images of vibrationally excited HC$_3$N emission (HC$_3$N$^*$) we reveal the presence of $8$ unresolved Super Hot Cores (SHCs) in the inner $160$\,pc of NGC\,253. Our LTE and non-LTE modelling of the HC$_3$N$^*$ emission indicate that SHCs have dust temperatures of $200-375$\,K, relatively high H$_2$ densities of $1-6\times 10^{6}$\,cm$^{-3}$ and high IR luminosities of $0.1-1\times 10^8$\,L$_\odot$. As expected from their short lived phase ($\sim 10^4$\,yr), all SHCs are associated with young Super Star Clusters (SSCs). We use the ratio of luminosities from the SHCs (protostar phase) and from the free-free emission (ZAMS star phase), to establish the evolutionary stage of the SSCs. The youngest SSCs, with the larges ratios, have ages of a few $10^4$\,yr (proto-SSCs) and the more evolved SSCs are likely between $10^5$ and $10^6$\,yr (ZAMS-SSCs). The different evolutionary stages of the SSCs are also supported by the radiative feedback from the UV radiation as traced by the HNCO/CS ratio, with this ratio being systematically higher in the young proto-SSCs than in the older ZAMS-SSCs. We also estimate the SFR and the SFE of the SSCs. The trend found in the estimated SFE ($\sim40\%$ for proto-SSCs and $>85\%$ for ZAMS-SSCs) and in the gas mass reservoir available for star formation, one order of magnitude higher for proto-SSCs, suggests that star formation is still going on in proto-SSCs. We also find that the most evolved SSCs are located, in projection, closer to the center of the galaxy than the younger proto-SSCs, indicating an inside-out SSC formation scenario.
\end{abstract}

\begin{keywords}
galaxies: individual: NGC\,253 -- galaxies: star clusters  -- galaxies: star formation -- galaxies: ISM -- galaxies: nuclei
\end{keywords}


\section{Introduction}
\label{sec:introduction}

 Starburst galaxies efficiently convert large amounts of gas and dust into stars in very short timescales,  from $10^{7-8}$\,yr  \citep[][]{Larson1978}.
 In these galaxies, a large fraction of the star formation is believed to be concentrated in relatively small regions in their nuclei, known as Super Star Clusters (SSCs). SSCs are compact star clusters, with sizes of $\approx 1$\,pc, massive ($\text{M}_* \gtrsim 10^5$\,$\text{M}_\odot$ ) and young (from a few to $100$\,Myr) \citep[][]{Whitmore1995, Beck2015}, and have been identified as probable progenitors of Globular Clusters \citep[GC,][]{Portegies2010}.
 Very likely, this extreme mode of star formation dominates in merging systems, and might be central in objects with a Star Formation Rate (SFR) in excess of $\sim 100$\,$\text{M}_\odot/\text{yr}$ at high redshift, when galaxy merging occurred more frequently \citep[][]{Clark2005}.
 Understanding the formation and evolution of SSCs in nearby galaxies is crucial to establish the conditions triggering the emergence of the starburst, to understand the processes that lead to cluster formation, and also to evaluate the effect of their associated radiative and kinematic feedback on the evolution of galaxies.
 
 So far, most of the studies on SSCs have been carried out in the optical and near-IR,  detecting  relatively evolved SSCs that have already cleaned their environment. Evolved SSCs with moderate visual extinctions have been observed with the Hubble Space Telescope (HST) in a certain number of starburst galaxies and mergers \citep[see][for a review]{Whitmore1995, Whitmore2002, Beck2015}.  Unfortunately, the earliest phases of SSCs formation and their evolution are poorly known since they are still deeply embedded in the parental cloud, hidden behind large columns of dust that avoid their observation even in the mid-IR.

With the advent of ALMA, the earliest phases of the SSCs can be studied at wavelengths free from extinction, shedding light on their formation and early evolution.
Based on ALMA high angular resolution ($0.11^{\prime \prime}$) images of dust emission in the nearby starburst galaxy NGC\,253 \citep[$3.5$\,Mpc][]{Rekola2005}, \citet{Leroy2018} have identified $14$ compact condensations with sizes of $2-3$\,pc, gas masses of a few $10^5$\,$\text{M}_\odot$ and dust temperatures of $\sim50-70$\,K. \citet{Leroy2018} have proposed that these condensations represent the precursors of the SSCs observed in the optical and IR after the removal of the material left from their formation. 

In the Milky Way (MW), the earliest phase (a few $10^4$\,yr) of massive star formation in clusters (proto-clusters) is commonly recognized as very compact ($0.02-0.1$\,pc), hot ($200-300$\,K), and dense condensations ($\text{n}_{\text{H}_2} \approx 10^7 \ \text{cm}^{-3}$), known as Hot Cores - HCs \citep[][]{Garay1999, Kurtz2000, Hoare2007}.  The HCs, with luminosities of  $10^5-10^7 \ \text{L}_\odot$ , are heated by massive protostars deeply embedded in molecular clouds \citep[][]{Wood1989, Osorio1999}. HCs would be best observed in the mid-IR ($10 \ \mu\text{m} - 50 \ \mu\text{m}$), where most of the hot dust emission peaks, but unfortunately, they are hidden behind very large extinctions preventing their direct observation at these wavelengths. Fortunately, HCs contain a large variety of molecules whose rotational emission at radio wavelengths can be used to study the kinematics and the physical properties of their inner parts \citep{Rivilla2017}.

Among these molecules, cyanoacetylene (HC$_3$N) is an excellent tool to study the properties of the proto-clusters since: i) its abundance is enhanced by its evaporation from grain mantles, and ii) its vibrational levels $v_7$, $v_6$ and $v_5$ with energies $310.7$, $719.4$ and $959.2$\,K above the ground state, respectively, are excited by IR radiation in the $45$\,$\ \mu\text{m}$ to $15$\,$\mu\text{m}$ range. Thus, the emission from the rotational transitions in vibrationally excited states of HC$_3$N (hereafter HC$_3$N*) can be used to probe the high density hot material surrounding the protostars \citep[e.g.][]{deVicente2000, Martin-Pintado2005} unaffected by dust extinction. For these reasons, HC$_3$N* has been successfully used to study the physical and kinematic properties of proto-clusters in the MW \citep[][]{Goldsmith1982, Wyrowski1999, deVicente2000, deVicente2002}, in NGC\,4418 \citep[][]{Costagliola2010} and in Arp\,220 \citep{Martin2011}.

Using ALMA observations of NGC\,253, we study the HC$_3$N emission from the rotational transition $\text{J}=24-23$ at $218-219$\,GHz and $\text{J}=39-38$ at $354-355$\,GHz in the ground state $v=0$ and vibrational levels $v_7=1$, $v_7=2$ and $v_6=1$ in order to identify and study the properties of the forming SSCs in this galaxy. From within the SSCs, we have identified $8$ sources in HC$_3$N* emission, which seems to trace the phase where SSCs are dominated by protostars (hereafter proto-SSCs), just before massive stars ionize their surroundings.

\section{Data reduction}
\label{sec:data_reduction}

We have used data from the public ALMA Science Archive in order to detect and analyze the properties of the HC$_3$N emission from the nucleus of NGC\,253. For our HC$_3$N analysis of the rotational transitions from the ground state  $v=0$ ($v_0$, hereafter) and the $v_7$ ($v_7=1$ and $v_7=2$) and $v_6$ ($v_6=1$) vibrationally excited states, we have used the observations summarized in Table~\ref{tab:observations}. 
Other observations containing HC$_3$N emission were also inspected (HC$_3$N transitions are spaced every $\sim 9.1$\,GHz), but we used the observations listed in Table~\ref{tab:observations} because they had the best angular resolution and sensitivity at the moment. The somewhat different angular resolution between the observations will not impact the analysis since HC$_3$N$^*$ emission is very compact.

\begin{table}
  \caption{ALMA observations used for HC$_3$N lines. Assuming a distance of $3.5\pm0.2$\,Mpc to NGC\,253 \citep[][]{Rekola2005}, $1^{\prime \prime}$ corresponds to $17\pm 1$\,pc.}
  \centering
	\centering
    \setlength{\tabcolsep}{4pt}
	\label{tab:observations}
	\begin{tabular}{lccc} 
		\hline
		Project Code  & Frequency & Resolution & rms \\
         &  \scriptsize (GHz) &\scriptsize (arcsec) & \scriptsize (mJy) \\
		\hline
		2013.1.00191.S  	& 217.92 - 219.82 	& $0.19^{\prime \prime}\times0.29^{\prime \prime}$	&	0.12\\
        2013.1.00973.S		& 292.03 - 307.89	& $0.37^{\prime \prime}\times0.49^{\prime \prime}$	&	0.86\\
        2013.1.00735.S  	& 340.07 - 355.80 	& $0.30^{\prime \prime}\times0.25^{\prime \prime}$	&	0.75\\
		\hline
	\end{tabular}
\end{table}

The data reduction was carried with Common Astronomy Software Applications \citep[CASA,][]{McMullin2007} version 4.2.2\,. To image the central region of NGC\,253 we have used CASA's \texttt{clean} task with Briggs weighting for deconvolution, setting the \texttt{robust} parameter to $0.5$ (in order to obtain the best possible trade-off between resolution and sensitivity) and a velocity resolution of $5$\,km\,s$^{-1}$.
After reduction, we applied a primary beam correction. The achieved synthesized beam sizes are given in Table~\ref{tab:observations}.  Continuum maps were built from line-free channels in the UV-plane.
The resulting rms measured are listed in Table~\ref{tab:observations}.

The data cubes generated with CASA without continuum subtraction were exported to MADCUBA\footnote{Madrid Data Cube Analysis (MADCUBA) is a software developed in the Center of Astrobiology (Madrid) to visualize and analyze data cubes and single spectra \citep{MADCUBA2019}.  Website: \url{http://cab.inta-csic.es/madcuba/MADCUBA_IMAGEJ/ImageJMadcuba.html}}  for line identification and Local Thermodynamic Equilibrium (LTE) analysis. Due to the richness of the molecular emission and the large velocity gradients across the nucleus, UV-plane subtracted continuum was not applied since it did not provide flat spectral baselines over the whole field of view. Further polynomial baselines of order $1$ were fitted and subtracted to produce the final data cubes. The resulting rms measured from line-free channels in the spectra is $\sim 1.5$\,mJy\,beam$^{-1}$ for $217-220$\,GHz, $\sim 1.2$\,mJy\,beam$^{-1}$ for $292-307$\,GHz and $\sim 1.1$\,mJy\,beam$^{-1}$ for $340-356$\,GHz. 

\begin{table}
\begin{center}
\caption[]{\label{tab:coordinates}Coordinates for the forming SSCs in NGC\,253 derived from the HC$_3$N$^*$ $J=24-23$ emission map. Positions with no HC$_3$N detection, marked with $^*$, were taken from the peak intensity of the dust continuum map.}
\setlength{\tabcolsep}{20pt}
\resizebox{\linewidth}{!}{%
\begin{tabular}{ccc}
\hline \noalign {\smallskip}
 SSC &  RA & Dec 	\\
 &  $(\text{J}2000)$ & $(\text{J}2000)$ \\
\hline \noalign {\smallskip}
1	& $	00^{\text{h}}47^{\text{m}}32^{\text{s}}.8044	$ & $	-25^{\circ}17^{\prime}21.21^{\prime \prime}	$ \\
2	& $	00^{\text{h}}47^{\text{m}}32^{\text{s}}.8199	$ & $	-25^{\circ}17^{\prime}21.24^{\prime \prime}	$  \\
3	& $	00^{\text{h}}47^{\text{m}}32^{\text{s}}.8287	$ & $	-25^{\circ}17^{\prime}21.13^{\prime \prime}	$  \\
4	& $	00^{\text{h}}47^{\text{m}}32^{\text{s}}.9415	$ & $	-25^{\circ}17^{\prime}20.19^{\prime \prime}	$  \\
5	& $	00^{\text{h}}47^{\text{m}}32^{\text{s}}.9811	$ & $	-25^{\circ}17^{\prime}19.71^{\prime \prime}	$  \\
6$^*$	& $	00^{\text{h}}47^{\text{m}}33^{\text{s}}.0101	$ & $	-25^{\circ}17^{\prime}19.42^{\prime \prime}	$  \\
7$^*$	& $	00^{\text{h}}47^{\text{m}}33^{\text{s}}.0123	$ & $	-25^{\circ}17^{\prime}19.08^{\prime \prime}	$  \\
8	& $	00^{\text{h}}47^{\text{m}}33^{\text{s}}.1141	$ & $	-25^{\circ}17^{\prime}17.64^{\prime \prime}	$  \\
9	& $	00^{\text{h}}47^{\text{m}}33^{\text{s}}.1141	$ & $	-25^{\circ}17^{\prime}18.19^{\prime \prime}	$  \\
10	& $	00^{\text{h}}47^{\text{m}}33^{\text{s}}.1517	$ & $	-25^{\circ}17^{\prime}17.11^{\prime \prime}	$  \\
11	& $	00^{\text{h}}47^{\text{m}}33^{\text{s}}.1671	$ & $	-25^{\circ}17^{\prime}17.44^{\prime \prime}	$  \\
12	& $	00^{\text{h}}47^{\text{m}}33^{\text{s}}.1760	$ & $	-25^{\circ}17^{\prime}17.20^{\prime \prime}	$  \\
13	& $	00^{\text{h}}47^{\text{m}}33^{\text{s}}.1959	$ & $	-25^{\circ}17^{\prime}16.69^{\prime \prime}	$  \\
14	& $	00^{\text{h}}47^{\text{m}}33^{\text{s}}.2932	$ & $	-25^{\circ}17^{\prime}15.52^{\prime \prime}	$  \\
\hline
\end{tabular}}
\end{center}
\end{table}

\section{Analysis}
\label{ref:Analysis}

\begin{figure*}
\centering
    \includegraphics[width=\linewidth]{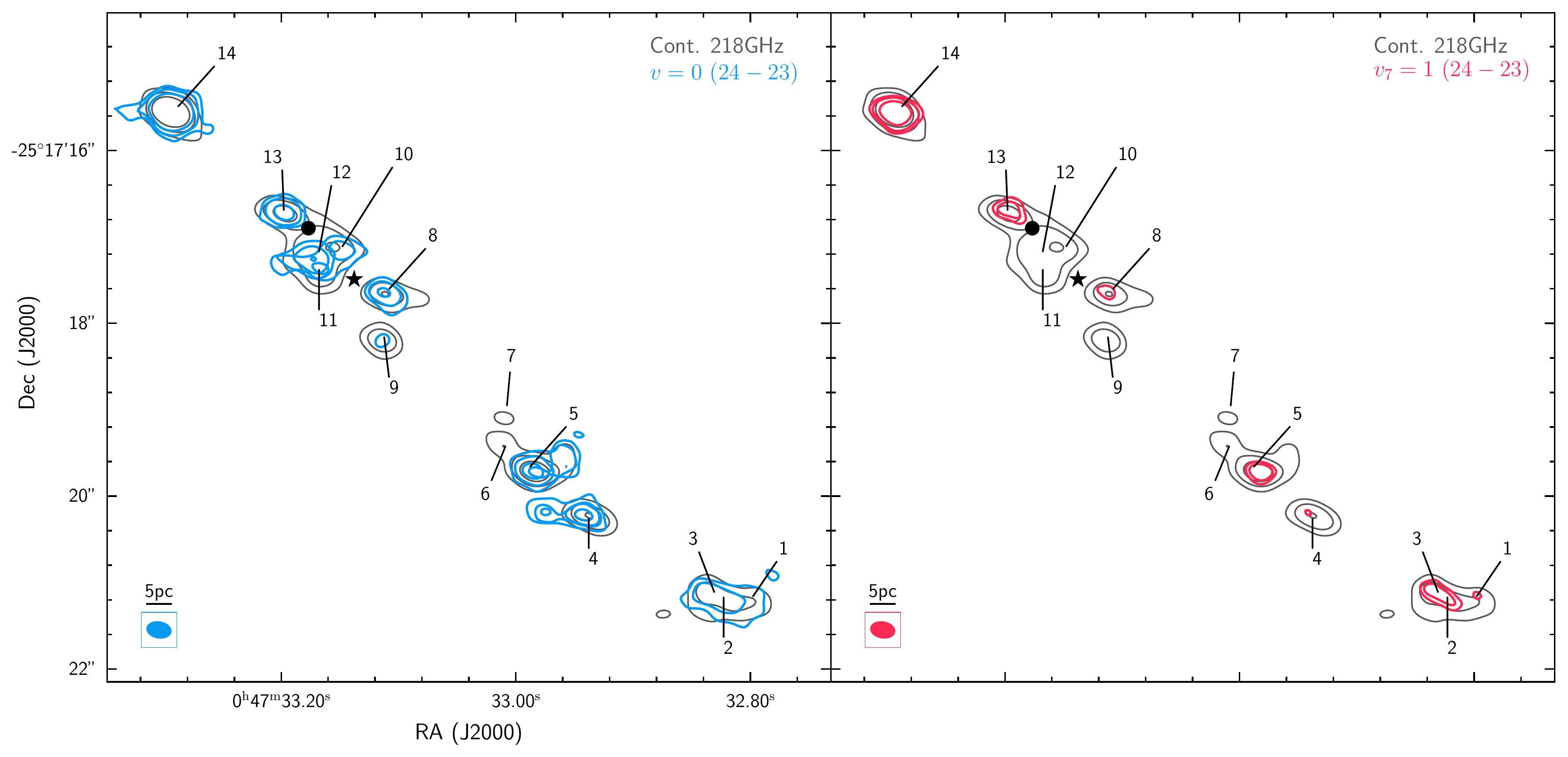}
  \caption{Grey contours show the $218$\,GHz continuum emission. On the left panel the HC$_3$N $v=0$ $J=24-23$ integrated line emission contours are overlaid in blue. On the right panel the HC$_3$N$^*$ $v_7=1$ $J=24-23$ integrated line emission contours are overlaid in red. All contours represent the $3\sigma$, $7\sigma$ and $15\sigma$ levels. The black circle indicates the position of TH2, $(\alpha(J2000), \delta(J2000))=(00^{\text{h}}47^{\text{m}}33^{\text{s}}.179, -25^{\circ}17^{\prime}17.13^{\prime \prime})$, \citep[strongest compact radio source][]{Ulvestad1997} and the black star the kinematic center as indicated by \citet{MullerSanchez2010}. The beam size ($0.19^{\prime \prime}\times 0.29^{\prime \prime}$) is indicated on the lower left corner of the panels.}
  \label{fig:hc3nv7_both}
\end{figure*}

 Following \citet{Leroy2018} notation from $350$\,GHz observations, we have identified the same $14$ clumps
from the peaks of either the HC$_3$N$^*$
and/or the $218$\,GHz continuum emission (Table~\ref{tab:coordinates}). 
Figure~\ref{fig:hc3nv7_both} shows the spatial distribution of the  $v=0$\,$J=24-23$ (in blue) and $v_7=1$\,$J=24-23$ (in red) integrated line intensities superimposed on the continuum emission (in grey) at $218$\,GHz. 
The HC$_3$N$^*$ high-$J$ ($\geqslant 24$) lines
trace the high density and hot ($\text{T}_{\text{ex}} > 200$\,K, see Sec.~\ref{subsec:SLIM}
condensations in the inner $100$\,pc of the nucleus of NGC\,253. 
The positions derived from the HC$_3$N$^*$ map (see Table~\ref{tab:coordinates}) coincide with the dust condensations observed in continuum emission within the uncertainties. These sources are unresolved by the beam, indicating sizes of $<0.1^{\prime \prime}$ ($<1.7$\,pc),  smaller than the dust continuum condensations ($2-3$\,pc) measured by \citet{Leroy2018}.

Further spectral analysis of each source was carried with the MADCUBA's tool SLIM (Spectral Line Identification and Modelling). A sample of spectra is shown in Fig.~\ref{fig:spec_SLIM} for clump $14$, the most luminous forming SSC. With SLIM we performed the line identification and the LTE analysis using the publicly available molecular catalogs CDMS\footnote{\url{http://www.astro.uni-koeln.de/cgi-bin/cdmssearch}} \citep{CDMS1,CDMS2} and JPL\footnote{\url{http://spec.jpl.nasa.gov/ftp/pub/catalog/catform.html}} \citep{JPL}.
We identified the HC$_3$N   $J=24-23$ and $J=39-38$ rotational transitions from the ground state $v=0$ and the $v_7=1$ and $v_6=1$ vibrationally excited levels. In addition we also measured the $J=24-23$ and $J=32-31$ HC$_3$N  rotational transitions for the $v_7=2$ vibrational state.

Table~\ref{tab:table_observ} lists the spectroscopic parameters and line fluxes (or upper limits) of the detected lines used for the analysis.
The detection criterion is an integrated intensity above the $3\sigma$ level over the  full linewidth  as derived from HC$_3$N $v=0$ lines (CS for sources with no HC$_3$N), with non-detections  indicated as upper limits in Table~\ref{tab:table_observ}.
From the $14$ dust condensations, $8$ are detected in HC$_3$N$^*$ emission and we will refer to them as Super Hot Cores (SHCs, see Section~\ref{subsec:SLIM} for details).
Among the remaining $6$ condensations, $4$ exhibit HC$_3$N $v=0$ emission (clumps $9$, $10$, $11$ and $12$) but clumps $6$ and $7$, remain undetected even in these HC$_3$N $v=0$ lines (see Table~\ref{tab:table_observ} and Fig.~\ref{fig:hc3nv7_both}).

The observed transitions involve energy levels that range from $121$\,K for the $v=0$ $J=24-23$ transition to $1042$\,K for the $v_6=1$ $J=39-38$ transition. The detection of high-$J$ transitions ($J\geqslant 24$) within the vibrationally excited states reveals that these sources are characterized by high excitation, which requires high temperatures and densities and/or, more likely (see below), mid-IR radiation emitted by warm dust.

\begin{table*}
\begin{center}
\caption[]{\label{tab:table_observ}MADCUBA fitted emission in Jy\,beam$^{-1}$\,km\,s$^{-1}$ for HC$_3$N and HC$_3$N$^*$ lines used for the analysis. The table also lists the transition frequency in GHz, the $J$ levels implied and the energy of the lower level ($E_{LO}$).  
Transitions with same $J$ numbers from a same vibrational level but different frequency come from the $l$-splitting of the vibrational levels due to the interaction between the angular momentum of the vibrationally excited states and the rotational angular momentum of the molecule \citep[][]{Goldsmith1983}.}
\setlength{\tabcolsep}{4pt}
\begin{tabular}{ccccccccccccccc}
\hline \noalign {\smallskip}
	&	 V$_{\text{LSR}}$ &	 FWHM	&	v=0	&	v=0	&	$v_7=1$	&	$v_7=1\textsuperscript{a}$	&	$v_6=1$	&	$v_6=1$	&	$v_7=2$	&	$v_7=2$	&	$v_7=2$	&	$v_7=2$	&	$v_7=2$	&	$v_7=2$	\\
 \scriptsize $J\rightarrow J-1$ & $(\text{km}$\,$\text{s}^{-1})$	& $(\text{km}$\,$\text{s}^{-1})$ &\scriptsize	24-23	& \scriptsize	39-38	& \scriptsize	24-23	&	 \scriptsize 39-38	& \scriptsize	24-23	& \scriptsize	39-38	& \scriptsize	24-23	&\scriptsize	24-23	& \scriptsize	24-23	& \scriptsize	32-31	& \scriptsize	32-31	& \scriptsize	32-31	\\ 
\scriptsize $l$-doubling & & & & & \scriptsize $(+1,-1)$ & \scriptsize $(+1,-1)$ & \scriptsize $(-1,+1)$ & \scriptsize $(+1,-1)$ & \scriptsize $(+2,-2)$ &  \scriptsize $(-2,+2)$ & \scriptsize $(0,0)$ & \scriptsize $(+2,-2)$ & \scriptsize $(-2,+2)$ & \scriptsize $(0,0)$ \\ 
\hline \noalign {\smallskip}
 \scriptsize $\nu$\,(GHz)	& 	& &\scriptsize 218.32	& \scriptsize	354.70	& \scriptsize	219.17	&	\scriptsize 355.57	& \scriptsize	218.68	& \scriptsize	355.28	& \scriptsize	219.74	& \scriptsize	219.71	& \scriptsize	219.68	& \scriptsize	292.99	& \scriptsize	\scriptsize 292.91	& \scriptsize	292.83	\\

\scriptsize $E_{LO}$\,(K) &	& & \scriptsize	120.50	& \scriptsize	323.48	& \scriptsize	441.81	&	 \scriptsize 645.11	& \scriptsize	838.35	& \scriptsize	1041.67	& \scriptsize	766.22	& \scriptsize	766.21	& \scriptsize	762.93	& \scriptsize	862.89	& \scriptsize	862.86	& \scriptsize	859.56	\\
\hline \noalign {\smallskip}
1		&	 $	310	\pm	1	$ 	&	 $	22	\pm	4	$ 	&	 $	0.18	$ 	&	 $	0.13	$ 	&	 $	0.07	$ 	&	 $	0.10	$ 	&	 $	\leqslant 0.05	$ 	&	 $	0.06	$ 	&	 $	-	$ 	&	 $	-	$ 	&	 $	-	$ 	&	 $	-	$ 	&	 $	-	$ 	&	 $	-	$\\
2		&	 $	307	\pm	1	$ 	&	 $	25	\pm	2	$ 	&	 $	0.33	$ 	&	 $	0.23	$ 	&	 $	0.14	$ 	&	 $	0.31	$ 	&	 $	0.08	$ 	&	 $	0.19	$ 	&	 $	\leqslant 0.05	$ 	&	 $	\leqslant 0.05	$ 	&	 $	\leqslant 0.05	$ 	&	 $	0.08	$ 	&	 $	0.08	$ 	&	 $	0.08	$\\
3		&	 $	299	\pm	1	$ 	&	 $	28	\pm	2	$ 	&	 $	0.38	$ 	&	 $	0.28	$ 	&	 $	0.17	$ 	&	 $	0.22	$ 	&	 $	0.12	$ 	&	 $	0.11	$ 	&	 $	\leqslant 0.05	$ 	&	 $	\leqslant 0.05	$ 	&	 $	\leqslant 0.05	$ 	&	 $	0.09	$ 	&	 $	0.11	$ 	&	 $	0.11	$\\
4		&	 $	253	\pm	1	$ 	&	 $	25	\pm	3	$ 	&	 $	0.37	$ 	&	 $	0.28	$ 	&	 $	0.12	$ 	&	 $	0.15	$ 	&	 $	0.08	$ 	&	 $	0.10	$ 	&	 $	\leqslant 0.05	$ 	&	 $	\leqslant 0.05	$ 	&	 $	\leqslant 0.05	$ 	&	 $	0.09	$ 	&	 $	0.08	$ 	&	 $	0.06	$\\
5		&	 $	212	\pm	1	$ 	&	 $	43	\pm	2	$ 	&	 $	0.76	$ 	&	 $	0.81	$ 	&	 $	0.21	$ 	&	 $	0.29	$ 	&	 $	0.13	$ 	&	 $	0.17	$ 	&	 $	\leqslant 0.07	$ 	&	 $	\leqslant 0.07	$ 	&	 $	\leqslant 0.07	$ 	&	 $	0.21	$ 	&	 $	0.19	$ 	&	 $	0.22	$\\
6		&	 $	219	\pm	2\textsuperscript{b}	$ 	&	 $	38\textsuperscript{b}$           	&	$\leqslant 0.06$ 	&	$\leqslant0.05$	&	 $	-	    $ 	&	 $	-	    $ 	&	 $	-	$ 	&	 $	-	$ 	&	 $	-	$ 	&	 $	-	$ 	&	 $	-	$ 	&	 $	-	$ 	&	 $	-	$ 	&	 $	-	$\\
7		&	 $	252	\pm	2\textsuperscript{b}	$ 	&	 $	32\textsuperscript{b}$           	&	$\leqslant 0.06$ 	&	$\leqslant0.05$	&	 $	-	    $ 	&	 $	-	    $ 	&	 $	-	$ 	&	 $	-	$ 	&	 $	-	$ 	&	 $	-	$ 	&	 $	-	$ 	&	 $	-	$ 	&	 $	-	$ 	&	 $	-	$\\
8		&	 $	302	\pm	2	$ 	&	 $	26	\pm	3	$ 	&	 $	0.34	$ 	&	 $	0.35	$ 	&	 $	0.14	$ 	&	 $	0.21	$ 	&	 $	\leqslant 0.05	$ 	&	 $	0.09	$ 	&	 $	-	$ 	&	 $	-	$ 	&	 $	-	$ 	&	 $	-	$ 	&	 $	-	$ 	&	 $	-	$\\
9		&	 $	165	\pm	2	$ 	&	 $	36	\pm	4	$ 	&	 $	0.12	$ 	&	 $	0.08	$ 	&	$\leqslant0.06$	&	 $	\leqslant 0.05	$ 	&	 $	-	$ 	&	 $	-	$ 	&	 $	-	$ 	&	 $	-	$ 	&	 $	-	$ 	&	 $	-	$ 	&	 $	-	$ 	&	 $	-	$\\
10		&	 $	274	\pm	1	$ 	&	 $	25	\pm	3	$ 	&	 $	0.20	$ 	&	 $	0.33\textsuperscript{c}	$ 	&	$\leqslant0.05$	&	 $	\leqslant 0.04	$ 	&	 $	-	$ 	&	 $	-	$ 	&	 $	-	$ 	&	 $	-	$ 	&	 $	-	$ 	&	 $	-	$ 	&	 $	-	$ 	&	 $	-	$\\
11		&	 $	141	\pm	1	$ 	&	 $	41	\pm	2	$ 	&	 $	0.28	$ 	&	 $	0.32	$ 	&	$\leqslant0.07$	&	 $	0.16	$ 	&	 $	-	$ 	&	 $	-	$ 	&	 $	-	$ 	&	 $	-	$ 	&	 $	-	$ 	&	 $	-	$ 	&	 $	-	$ 	&	 $	-	$\\
12		&	 $	150	\pm	1	$ 	&	 $	46	\pm	3	$ 	&	 $	0.36	$ 	&	 $	0.18	$ 	&	$\leqslant0.07$	&	 $	\leqslant 0.06	$ 	&	 $	-	$ 	&	 $	-	$ 	&	 $	-	$ 	&	 $	-	$ 	&	 $	-	$ 	&	 $	-	$ 	&	 $	-	$ 	&	 $	-	$\\
13		&	 $	250 \pm	1	$ 	&	 $	31	\pm	1	$ 	&	 $	0.70	$ 	&	 $	0.71	$ 	&	 $	0.36	$ 	&	 $	0.39	$ 	&	 $	0.10	$ 	&	 $	0.18	$ 	&	 $	\leqslant 0.06	$ 	&	 $	\leqslant 0.06	$ 	&	 $	\leqslant 0.06	$ 	&	 $	0.09	$ 	&	 $	0.08	$ 	&	 $	0.10	$\\
14		&	 $	201	\pm	6	$ 	&	 $	51	\pm	14 $ 	&	 $	2.74    $ 	&	 $	3.56    $ 	&	 $	1.12	$ 	&	 $	1.59	$ 	&	 $	0.45	$ 	&	 $	0.78	$ 	&	 $	0.41	$ 	&	 $	0.40	$ 	&	 $	0.40	$ 	&	 $	0.69	$ 	&	 $	0.69	$ 	&	 $	0.71	$\\
\hline \noalign {\smallskip}
  
\end{tabular}
\begin{tablenotes}
      \small
      \item \textsuperscript{a} This transition is contaminated with the $v_6=1$ $J= 39-38$\,$(-1,+1)$ transition. The listed values have been corrected from this contamination using MADCUBA.
      \item \textsuperscript{b} Velocities and linewidths for sources with upper-limits in HC$_3$N $v=0$ were taken from CS and C$^{18}$O lines.
      \item \textsuperscript{c} Strongly blended with HCN.
    \end{tablenotes}
\end{center}
\end{table*}

\subsection{Radiative and Collisional Excitation of HC$_3$N}
\label{subsec:hc3n_excitation}

HC$_3$N is a linear molecule with seven vibrational modes, four stretching modes ($v_1$, $v_2$, $v_3$, $v_4$) and three bending modes ($v_5$, $v_6$, $v_7$) \citep[][]{Uyemura1982, Wyrowski1999}. It is an excellent probe of the physical properties of hot and dense regions in the MW (i.e. Hot Cores) where  massive star formation is taking place. In the warm regions where the gas is shielded from the UV radiation, the abundance of HC$_3$N is expected to increase due to the evaporation of this molecules from grain mantles. Furthermore, its high dipole moment \citep[$3.73$\,Debye,][]{deLeon1985}
traces high densities of  $n_{\text{H}_2} > 10^5$\,cm$^{-3}$. The vibrational levels of HC$_3$N$^*$ with energies ranging between $\sim 300$\,K and $\sim 1000$\,K  above the ground state for the bending modes and above $3000$\,K for the stretching modes are predominantly excited by warm $> 200$\,K mid-IR radiation. The $v_7 = 1$ and $v_6 = 1$ states can be excited via absorption of $45$ and $20$\,$\mu$m photons, respectively, and could also be pumped via collisions with H$_2$ in hot and dense regions; however, the latter mechanism is restricted to small regions while the former is expected to be more efficient at the spatial scales probed by our observations. Since the Spectral Energy Distribution (SED) of the dust emission in regions with hidden massive star formation usually peaks in the $10-60$\,$\mu$m region, it is expected that bending modes will be more easily excited than stretching modes, which require IR radiation at $5$\,$\mu$m.

As a consequence of the high column densities, the extinction in star forming regions is very high, preventing the direct observation of the hot dust in the mid-IR. However, the HC$_3$N rotational transitions from its ground and vibrationally excited states are emitted in the (sub)millimeter range which is basically unaffected by dust extinction \citep[even in extremely obscured objects like the nuclei of the ultraluminous IR galaxy Arp\,220,][]{Barcos2015, Martin2016}, allowing to probe deeply embedded sources. Therefore, measuring multiple rotational transitions from different vibrational states of HC$_3$N provides a unique tool to  infer their physical properties, their thermal and density structures, and the kinematics of the material heated by the protostars. Since the continuum optical depth in the mid-IR is expected to be high, the HC$_3$N molecules will be bathed by a blackbody at the local T$_\text{dust}$, and the upper vibrational levels will be populated accordingly.

The typical densities of the HCs in the MW (few $10^6-10^7$\,cm$^{-3}$) \citep[]{Wyrowski1999, deVicente2000} are usually much smaller than the critical densities ($n_{\text{cr}}$) required to collisionally excite the vibrational levels from the ground state. Values for $n_{\text{cr}}$ at $T=300$\,K are of $4\times 10^8$ and $3 \times 10^{11}$\,cm$^{-3}$ for the excitation of the $v_7=1$ and $v_6=1$ states , respectively \citep[][]{Wyrowski1999}, indicating that the excitation of the vibrational levels is usually dominated by radiation pumping in the mid-IR. Therefore, the detection of the rotational transitions from vibrationally excited levels can be used to infer the temperature of the warm dust. 
By contrast, collisions with H$_2$ may dominate the excitation of the rotational levels within the $v=0$, $v_6=1$, and $v_7=1$ vibrational states; the critical densities at $300$\,K to excite the $J=24-23$ and $J=39-38$ transitions are of $2 \times 10^{6}$ and $5 \times 10^{6}$\,cm$^{-3}$, respectively. While radiative pumping of the excited vibrational states and subsequent relaxation can potentially contribute to the rotational excitation, direct excitation through collisions will efficiently populate the rotational ladder within $v=0$, from which the excited vibrational states will be radiatively pumped. In summary, one expects a combination of collisional excitation of the rotational transitions within the  vibrational state of  HC$_3$N and radiative pumping for the the vibrational excitation.

\section{Results}
\label{sec:results}

\subsection{LTE modelling. Rotational and vibrational temperatures}
\label{subsec:LTE}

As discussed in the previous section, the excitation of the HC$_3$N lines is dominated by different mechanisms: collisional for rotational transitions and IR pumping for vibrational transitions. To establish if the excitation of HC$_3$N is described by the Local Thermodynamic Equilibrium (LTE) with a single excitation temperature we have used two LTE analysis: the rotational diagram and  the MADCUBA SLIM  tool \citep{MADCUBA2019}. On the one hand, the rotational diagrams simply assume optically thin emission. On the other hand, SLIM includes line optical depth effects by fitting not only the line ratios but also the absolute line fluxes and profiles with an assumed size for the source. For both analysis, we have first combined the relative intensities and line profiles of a given rotational transition  ($J= 24-23$, $J= 32-31$ or $J= 39-38$) arising from the ground state and the different vibrationally excited states to derive the excitation temperature between vibrational levels (hereafter the vibrational temperature, $T_{\text{vib}}$). The lines used to derive $T_{\text{vib}}$ cover a wide range of lower level energies, from $120$\,K to $1041$\,K for sources with detections of the $v_6=1$ lines; and to $645$\,K for sources with detections of only the $v_7=1$ lines.  Then, we have combined all the rotational  transitions arising from the same vibrationally states (ground state, $v_7=1$, $v_7=2$ and $v_6=1$) to derive the excitation temperature of the rotational levels within  the different vibrational states (hereafter the rotational temperature, $T_{\text{rot}}$).

\subsubsection{Rotational diagrams}
\label{subsec:Rotational}

\begin{figure}
\centering
    \includegraphics[width=\linewidth]{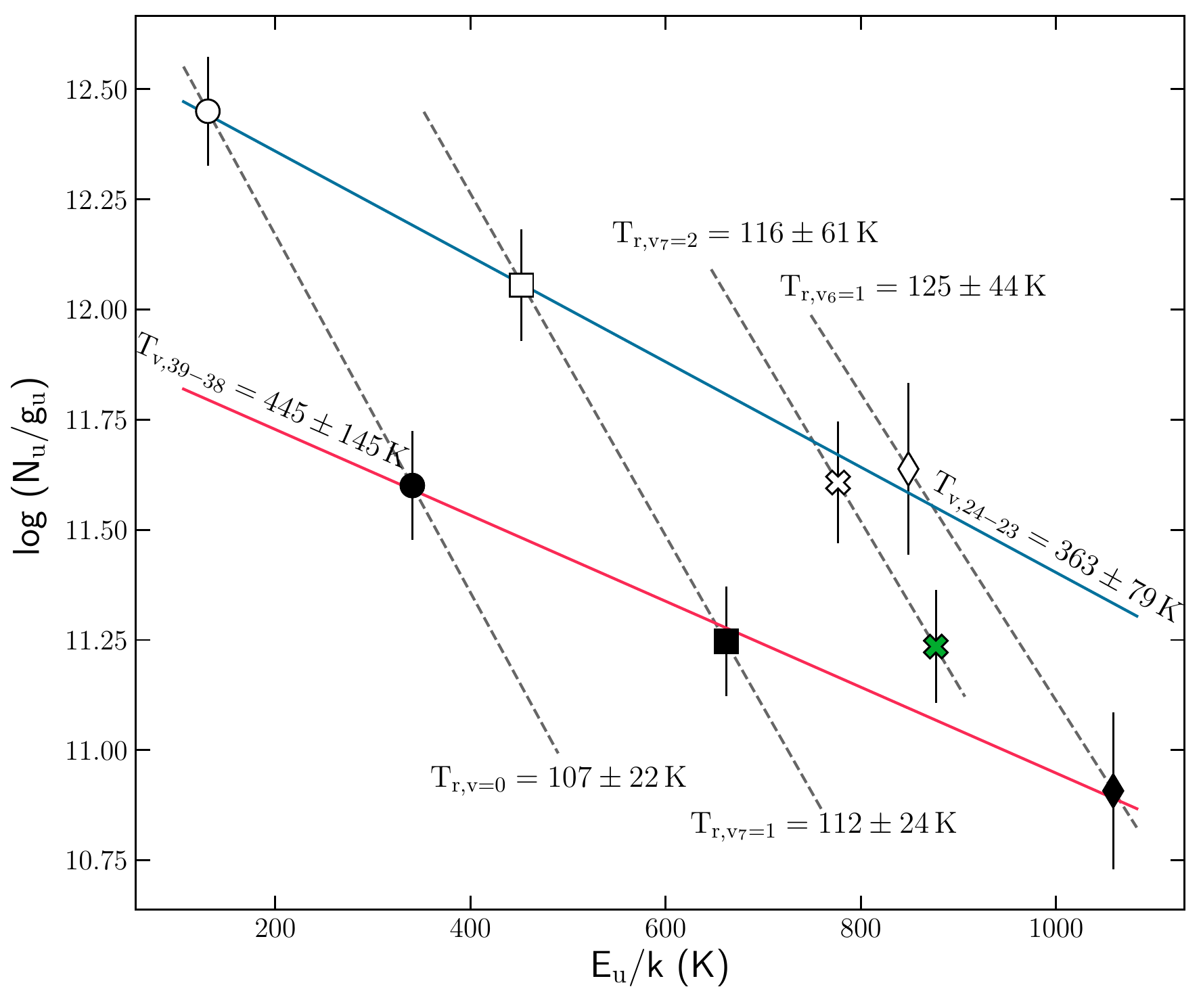}
  \caption{Rotational diagram derived from the line intensities of HC$_3$N$^*$ for source $14$. 
  The data from the $J = 24-23$ and $J = 39-38$ transition are indicated by empty and filled markers, respectively. The green colored cross represents the $J = 32-31$ transition from $v_7=2$. The population levels arising from different vibrational states $v=0$, $v_7=1$, $v_7=2$  and $v_6=1$ are represented as circles, squares, crosses and diamonds. 
  The blue and red solid lines represent the fit to the $J= 24-23$ and the $J= 39-28$ transitions, respectively. The dotted lines represent the fit to same vibrational state transitions. The temperatures derived from each fit are indicated.}
	\label{fig:rotdiag}
\end{figure}

Figure~\ref{fig:rotdiag} shows the LTE results for SHC14, the most prominent condensation, by combining all detected rotational lines in the different vibrational states to infer both $T_{\text{rot}}$ and $T_{\text{vib}}$. 
The rotational diagram (Fig.~\ref{fig:rotdiag}) clearly illustrates the presence of two different excitation temperatures, $T_{\text{vib}}$ of $363\pm79$ and $445\pm145$\,K  shown in blue and red solid lines respectively, derived from the $J= 24-23$ and  $J= 39-38$ transitions in different vibrational levels, and the $T_{\text{rot}}$ of $107\pm22$, $112\pm24$, $116\pm61$ and $125\pm45$\,K  shown by dotted lines derived from the different rotational transitions arising from  the same vibrational state. The rotational diagram clearly illustrates that $T_{\text{vib}}$ and $T_{\text{rot}}$ have quite different values of $\sim400$ and $\sim110$\,K, respectively.

\subsubsection{SLIM analysis}
\label{subsec:SLIM}
In addition to the rotational diagram analysis, the observed HC$_3$N line profiles from the ground and vibrationally excited levels have been fitted using the SLIM tool to derive the physical properties of the sources with HC$_3$N emission. 
SLIM simulates all lines profiles and intensities emitted under LTE conditions including line optical depth effects for a given source size.
The SLIM LTE analysis considers, as free parameters, the column density ($N$) of HC$_3$N, the excitation temperature ($T_{\text{vib}}$ or $T_{\text{rot}}$ depending on the transitions considered), the radial velocity ($V_\text{LSR}$), the linewidth and the size of the emitting source (defined as FWHM, full width at half maximum). Since the HC$_3$N$^*$ sources are unresolved by our beam of $ 0.2^{\prime \prime}$, we have adopted an upper limit to their sizes of $0.1^{\prime \prime}$, i.e. half of the beam size.
Therefore the inferred column densities could be lower limits if the source sizes are substantially smaller than the assumed value, nonetheless the derived excitation temperatures remain independent of this choice.

\begin{figure*}
\centering
    \includegraphics[width=\linewidth]{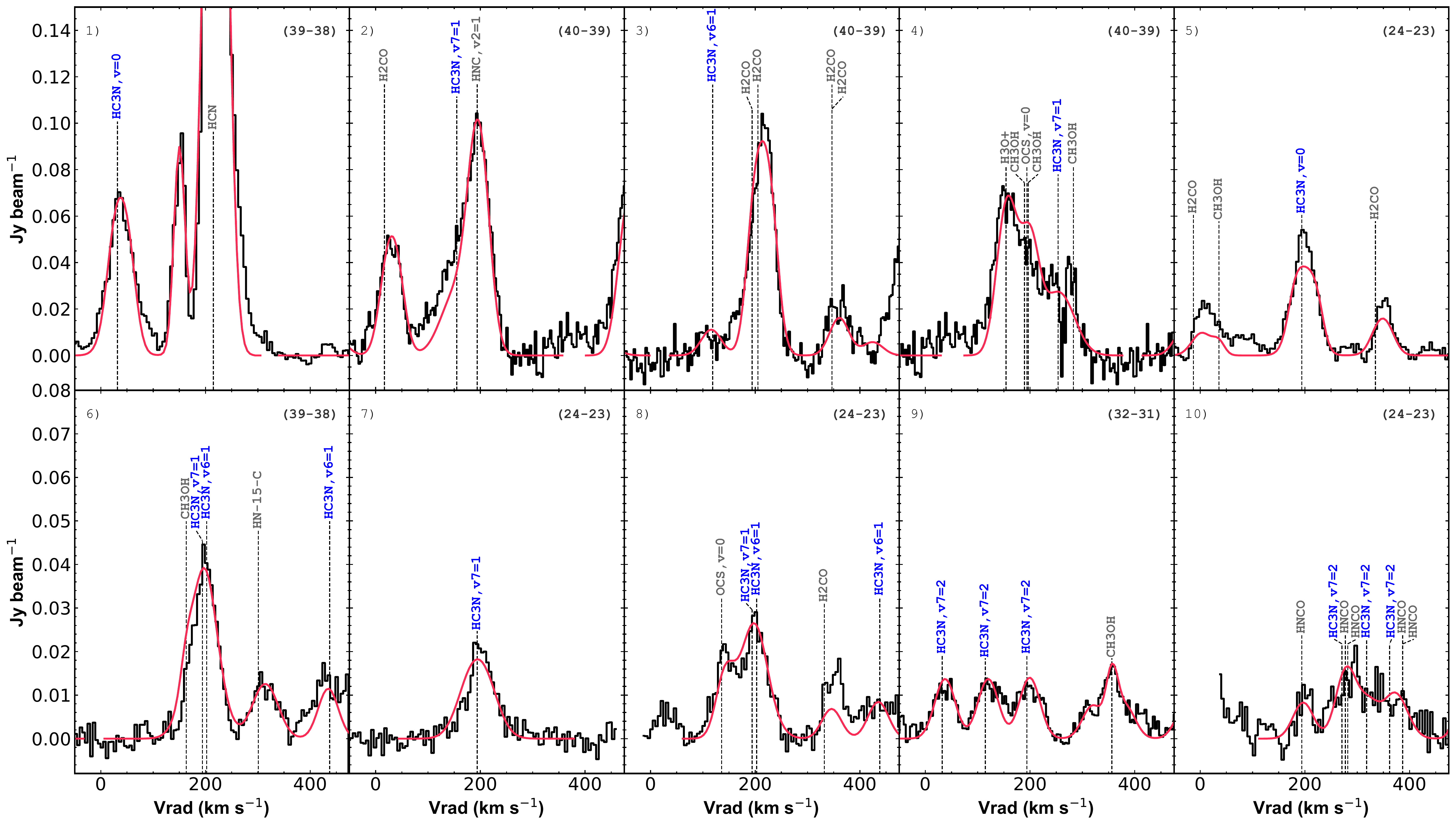}
      \caption{
      Observed spectra (black histograms) towards source $14$. HC$_3$N lines from the ground and vibrational states $v_7=1$, $v_7=2$, and $v_6=1$ are marked in blue. On the top-right corner of each panel the ($J$, $J-1$) of the HC$_3$N transitions present in that panel is indicated.  The red solid lines represent the fitted model from the LTE analysis obtained with MADCUBA.}
      \label{fig:spec_SLIM}
\end{figure*}

Figure~\ref{fig:spec_SLIM} shows the SLIM predictions of the line profiles for all  HC$_3$N$^*$ lines detected in SHC14 in red solid line superimposed on the observed spectra.  We have used the  AUTOFIT tool in SLIM, which performs a non-linear least squared fit of the LTE line profiles to the data using the Levenberg-Marquartd algorithm.  Since SLIM uses the partition function given by the CDMS catalog, which only uses the ground vibrational state to derive the partition function, we have corrected the estimated column densities by calculating the total partition function, including all rotational levels inside the ground state and vibrationally excited states ($v_7=1$, $v_6=1$ and $v_7=2$).  The corrected column densities and temperatures derived  from our LTE SLIM model fitting for all sources are summarized in Table~\ref{tab:table_temp}. The inferred temperatures are in general agreement with those derived via the rotational diagram method.

The inferred $T_{\text{vib}}$ by the two methods (SLIM and rotational diagram) are high, ranging from $\sim 216$\,K to $\sim 393$\,K for sources with at least one of the $v_7=1$ lines detected above the $3\sigma$ level (sources $1$, $2$, $3$, $4$, $5$, $8$, $13$ and $14$). 
It is interesting to note that the $T_{\text{rot}}$ within a given vibrational level does not vary significantly among different sources. However, there is a trend for $T_{\text{rot}}$ to increase with the energy of the vibrational level. The average rotational temperatures are: $T_{\text{rot}}= 91$\,K for $v=0$, $T_{\text{rot}}= 152$\,K for $v_7=1$  and  $T_{\text{rot}}=206$\,K for $v_6=1$. This is clear indication that in the case of collisional excitation of the rotational lines, the regions where the $v_6$ rotational transitions arise have larger H$_2$ densities (and kinetic temperatures) than those arising from the ground and the $v_7$ vibrational levels. This could imply the presence of density and temperature gradients in the structure of the SHCs \citep{deVicente2000}, as expected if they are associated with very recent star formation in the cloud.

We have also derived upper limits to $T_\text{vib}$ of $\sim 100-160$\,K for the massive star forming  regions $9$, $10$, $11$ and $12$, where no vibrationally excited lines were detected. 
As discussed in Section~\ref{sec:Discussion}, these sources (along with sources $6$ and $7$) likely represent a more evolved stage in the evolution of the formation of SSCs.

The differences found between the vibrational and rotational temperatures ($T_{\text{vib}} \gg T_{\text{rot}}$) and between the $T_{\text{rot}}$ for distinct vibrational states (see Figs.~\ref{fig:spec_SLIM} and \ref{fig:rotdiag} and Table~\ref{tab:table_temp}), clearly suggests that the HC$_3$N$^*$ excitation is not in LTE as expected when the H$_2$ density is not high enough to thermalize the population of the rotational levels. 

\begin{table}
\begin{center}
\caption[]{\label{tab:table_temp}Derived parameters for the SSCs in NGC\,253 from the SLIM LTE modelling assuming a source size of $0.1^{\prime \prime}$. Column densities have been corrected by including the contribution from the vibrational states into the partition function. Column density errors are indicated in parenthesis. For SSCs without a T$_\text{rot}$ estimation for $v_7=1$, a fiducial T$_\text{rot}=100$\,K was assumed to get an HC$_3$N column density upper limit.}
\begin{threeparttable}
\setlength{\tabcolsep}{3pt}
\begin{tabular}{ccccccc}
\hline \noalign {\smallskip}
SSC &	 &	log N(HC$_3$N)$\textsuperscript{a}$ 			&	T$_\text{vib}$			& \multicolumn{3}{c}{T$_\text{rot}$} \\\cmidrule{5-7}
	& 	&	& &
v=0			&	$v_7=1$			&	$v_6=1$			\\
	&	&	(cm$^{-2}$)			&	(K)			&	(K)			&	(K)			&	(K)			\\
    \hline \noalign {\smallskip}
1		&	SHC	&	 $	 16.3		(15.4)	$ 	&	 $	216	\pm	15	$ 	&	 $	84	\pm	9	$ 	&	 $	132	\pm	3	$ 	&	 $	259	\pm	19	$ \\
2		&	SHC	&	 $	 16.4		(15.2)	$ 	&	 $	304	\pm	70	$ 	&	 $	69	\pm	6	$ 	&	 $	170	\pm	4	$ 	&	 $	227	\pm	5	$ \\
3		&	SHC	&	 $	 16.2		(15.9)	$ 	&	 $	337	\pm	50	$ 	&	 $	78	\pm	4	$ 	&	 $	162	\pm	36	$ 	&	 $	240	\pm	15	$ \\
4		&	SHC	&	 $	 16.4		(16.2)	$ 	&	 $	326	\pm	52	$ 	&	 $	93	\pm	6	$ 	&	 $	142	\pm	25	$ 	&	 $	193	\pm	13	$ \\
5		&	SHC	&	 $	 16.6		(15.5)	$ 	&	 $	269	\pm	22	$ 	&	 $	110	\pm	5	$ 	&	 $	160	\pm	35	$ 	&	 $	130	\pm	12	$ \\
6		&		&	 $	\leqslant 14.6		$ 	&	 $  -$              &	 $	100	$	 	&	 $	-			$ 	&	 $	-			$ \\
7		&		&	 $	\leqslant 14.6		$ 	&	 $	- $ 	        &	 $	100$		 	&	 $	-			$ 	&	 $	-			$ \\
8		&	SHC	&	 $	 16.1		(15.1)	$ 	&	 $	217	\pm	14	$ 	&	 $	102	\pm	16	$ 	&	 $	165	\pm	45	$ 	&	 $	\sim 181	$ \\
9		&		&	 $	 \leqslant 16.2		$ 	&	 $  \leqslant 126$ 	&	 $	\leqslant 88	$ 	&	 $	-			$ 	&	 $	-			$ \\
10		&		&	 $	 \leqslant 16.0		$ 	&	 $  \leqslant 132$ 	&	 $	\leqslant 84 	$ 	&	 $	-			$ 	&	 $	-			$ \\
11		&		&	 $	 \leqslant 15.9		$ 	&	 $  \leqslant 165$ 	&	 $	113	\pm	10	$ 	&	 $	\leqslant 138	$ 	&	 $	-			$ \\
12		&		&	 $	 \leqslant 16.1		$ 	&	 $  \leqslant 140$ 	&	 $	76	\pm	6	$ 	&	 $	-			$ 	&	 $	-			$ \\
13		&	SHC	&	 $	 16.8		(16.3)	$ 	&	 $	393	\pm	34	$ 	&	 $	104	\pm	4	$ 	&	 $	159	\pm	21	$ 	&	 $	206	\pm	8	$ \\
14		&	SHC	&	 $	 17.6		(16.6)	$ 	&	 $	312	\pm	20	$ 	&	 $	96	\pm	13	$ 	&	 $	139	\pm	12	$ 	&	 $	124	\pm	29	$ \\
\hline \noalign {\smallskip}
  
\end{tabular}
\begin{tablenotes}
      \small
      \item \textsuperscript{a} $\log{N}=a(b)$ represent $N=10^a \pm 10^b$.
\end{tablenotes}
\end{threeparttable}
\end{center}
\end{table}

\subsection{Non-LTE modelling}

To properly account for the different excitation mechanisms of the vibrational and rotational transitions of HC$_3$N, we have carried out  non-LTE radiative transfer modelling of HC$_3$N, considering both the effects of the mid-IR radiation from the warm dust and the collisional excitation. We have used the radiative transfer code described in \citet[][]{GA97, GA99} to calculate the statistical equilibrium populations arising from a uniform spherical cloud.

In our model, we have included the HC$_3$N rotational transitions up to $J=45$ in the  $v=0$, $v_7=1$ and $v_6=1$ vibrational states. Since collisional rates with H$_2$ for the transitions from the ground to vibrationally excited levels are not available, we have estimated them following the approach described by \citet[][]{Deguchi1979}, \citet[][]{Goldsmith1982} and \citet[][]{Wyrowski1999}.
We have also assumed the same collisional rates for the excitation of the rotational levels by para- and ortho-H$_2$ for the  $v=0$, $v_7=1$ and the $v_6=1$ states \citep[taken from][]{Faure2016}, and considered that there are no propensity rules for the $l$-type doubling: $C_{v_7,v_6}(J_{\text{up}},l_{\text{up}} \rightarrow J_{\text{low}},l_{\text{low}}) = 0.5 C_{v0} (J_{\text{up}}, J_{\text{low}})$. Here $C_{v_7,v_6}$ and $C_{v0}$ are  the collisional rates for a rotational transition in the vibrational level $v_7=1$ or $v_6=1$ and in the ground state, respectively.

For a given dust temperature ($T_\text{dust}$) and a dust column density, the model calculates the radiation field from the mid-IR to millimeter wavelengths within a uniform spherical cloud. This radiation field is used to radiatively pump the rotational levels in the $v_7=1$ and $v_6=1$ vibrationally excited states. 
The model also returns the SED of the dust emission (i.e a grey body) emerging from the spherical cloud, which is integrated to determine the total luminosity. In addition to the radiative excitation dominated by the dust, the model also calculates the collisional excitation of HC$_3$N for a given H$_2$ density assuming that the gas kinetic temperature is equal to the dust temperature.
This assumption is justified by the relatively large H$_2$ densities of $\sim 10^6$\,cm$^{-2}$ derived from our LTE modelling (see Section~\ref{sec:results}), for which both temperatures should be closely coupled, as seen in MW HCs \citep[][]{deVicente2000}.
For the line radiation transfer, the model assumes a Gaussian line profile with the linewidth as a free parameter. We have used the model to predict the HC$_3$N emission for the rotational lines from the $v_6=1$, $v_7=1$ and  $v=0$ vibrational states as a function of the H$_2$ density, the dust/kinetic temperature, the HC$_3$N column density and the dust column density parameterized by the dust opacity at $100$\,$\mu$m, $\tau_{100}$.  The $\tau_{100}$ can be transformed into the gas H$_2$ column density for a gas-to-dust ratio of $100$ by
\begin{center}
\begin{align}
    N(\text{H}_2) = 6.5 \times 10^{23} \tau_{100} \, (\text{cm}^{-2})
\end{align}
\end{center}
where the mass absorption coefficient of dust at $100$\,$\mu$m has been  taken to be $44.5$\,$\text{cm}^2\text{g}^{-1}$ \citep{GA2014} with a dust emissivity index of $1.6$.

\begin{figure*}
\centering
    \includegraphics[width=\linewidth]{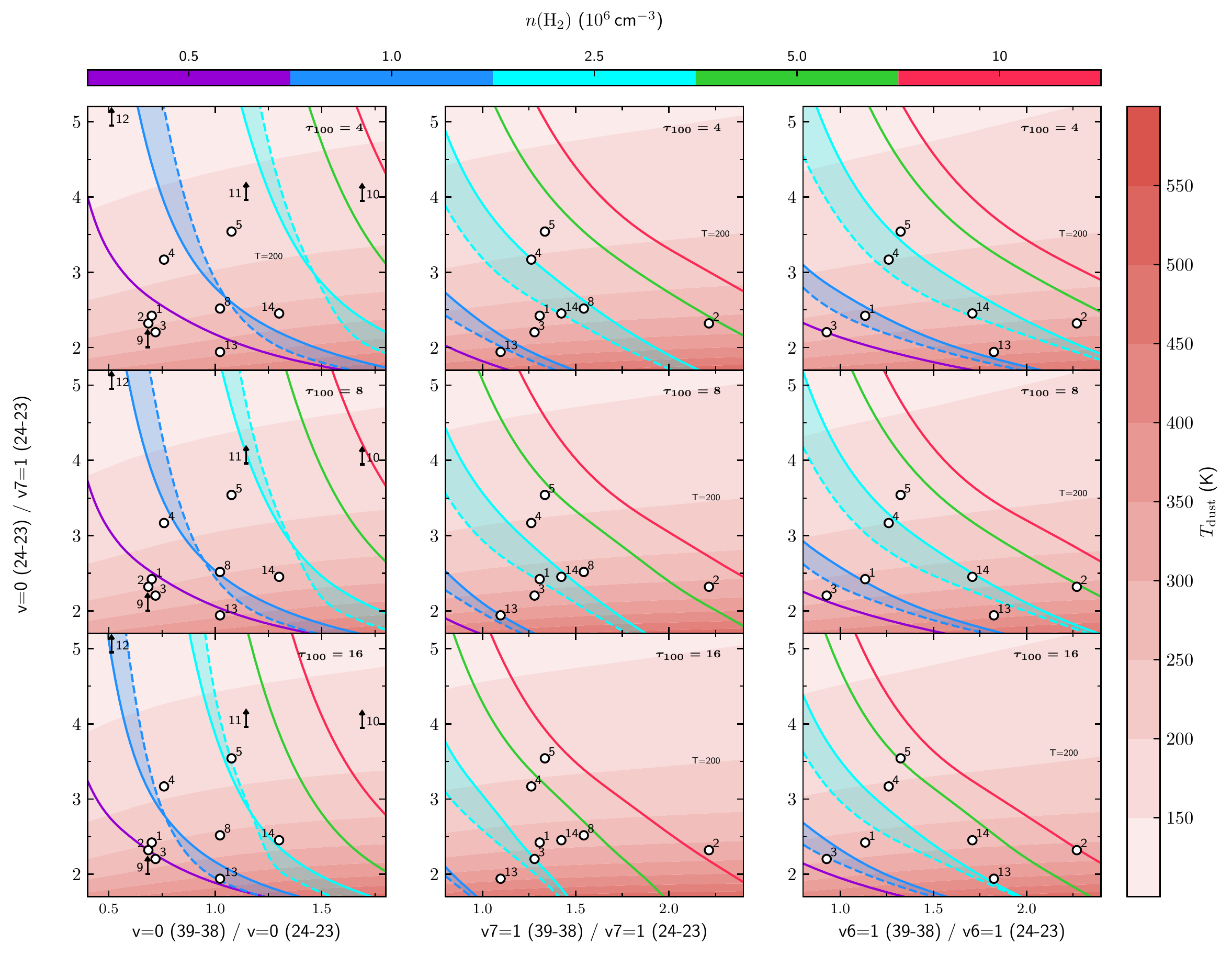}
      \caption{HC$_3$N non-LTE modelling results summarized as the line intensity ratio between the rotational transition $J=24-23$ from the vibrational state  $v=0$ and $v_7=1$ versus the line intensity ratio between the $J=39-38$ and the $J=24-23$ rotational transition from the ground state $v=0$ (left panels) and from the vibrational states $v_7=1$ (middle panels) and  $v_6=1$ (right panels). Red  contours indicate dust/kinetic temperatures, ranging from $100$\,K to $600$\,K. 
      Colored solid lines indicate the model dependence with density for an HC$_3$N column density of $5\times 10^{16}$\,cm$^{-2}$.
      The H$_2$ density corresponding to each color is indicated on the horizontal colorbar.
      The colored shaded regions show the effect of varying the HC$_3$N column density from $5\times 10^{16}$\,cm$^{-2}$ (solid lines) to $10^{17}$\,cm$^{-2}$ (dashed lines) on models with H$_2$ densities of $1\times 10^6$ (blue) and $2.5\times 10^6$\,cm${-3}$ (cyan).
      The top, center and bottom panels show the results for dust opacities  $\tau_{100}=4$, $8$ and $16$, respectively. These values translate into  $N(\text{H}_2) = 2.6\times 10^{24}$, $5.2\times 10^{24}$ and   $1.04\times 10^{25}$\,cm$^{-2}$ for the top, middle and bottom panels.  
      The observed ratios for the SHCs are shown with filled black white circles and  sources without the detection of $v_7=1$ lines are represented with black arrows only in the left panel. }
      \label{fig:ratios}
\end{figure*}

The predictions of our non-LTE modelling for a spherical cloud with uniform density and dust/kinetic temperature are summarized in Fig.~\ref{fig:ratios}.
This figure shows the predicted intensity ratio between the rotational transition, $J=24-23$, from the vibrational levels $v=0$ and $v_7=1$ (hereafter  $v_0/v_7$ ratio) plotted against the predicted intensity ratio between two rotational transitions $J=39-38$ and $J=24-23$ from the ground state $v=0$ ($v_0$ ratio) on the left panel, the vibrational state $v_7=1$ ($v_7$ ratio) on the middle panel and the vibrational state $v_6=1$ ($v_6$ ratio) on the right panel. 
The continuum optical depth at $100$\,$\mu$m, $\tau_{100}$, has been varied from $4$ in the upper panels to $16$ in the lower panels, covering the relevant H$_2$ gas column densities from $2.5 \times 10^{24}$ to $1.0 \times 10^{25}$\,cm$^{-2}$. 
The colored solid lines in all panels indicate the dependence of the line intensity ratios on  H$_2$ densities and the reddish contour levels show the dependence of the line ratios on the dust/kinetic temperature. 
For Figure~\ref{fig:ratios} a representative HC$_3$N column density of $5\times10^{16}$\,cm$^{-2}$ was assumed (solid lines), however, to show the dependency on the HC$_3$N column density (i.e. the HC$_3$N abundance), we have shaded the regions that cover a  column density varying from $5\times 10^{16}$\,cm$^{-2}$ (solid lines) to $1\times 10^{17}$\,cm$^{-2}$ (dashed lines)  for two different H$_2$ densities: $1.0\times 10^6$ and $2.5\times 10^6$\,cm$^{-3}$ (blue and cyan shaded regions).
These  HC$_3$N column densities ($5\times10^{16}$\,cm$^{-2}$ and $1\times10^{17}$\,cm$^{-2}$) translate into HC$_3$N abundances ranging from $X$(HC$_3$N)$=4\times10^{-9}$ to $2\times10^{-8}$.

Figure~\ref{fig:ratios} clearly shows the expected trends for the different line ratios. The  vibrational $v_0/v_7$ ratio is extremely sensitive to the dust temperature with the iso-contours of $T_\text{dust}$ running basically horizontally. The $v_0$, $v_7$ and $v_6$  line ratios show a nearly linear dependence with $n_{\text{H}_2}$ density (colored lines) with iso-density lines running from upper-left to the bottom-right, with density increasing from  left to right (H$_2$ densities corresponding to each color are indicated on the horizontal color bar).
On the other hand, the line ratios have a weak dependence on the HC$_3$N abundance/column density as shown with the dashed lines, although this dependence increases as the density increases.
Finally, the dust opacity has a weak effect on the derived dust temperature and a moderate effect on the derived H$_2$ densities. For a given vibrational ratio $v_0/v_7$, an increase of the dust opacity by a factor of $4$ requires increasing the dust temperature by just only a factor of $\sim 1.2$ and the H$_2$ density by less than a factor of $2$.  

In Figure~\ref{fig:ratios} we have included, as open circles, the observed ratios from all sources (labelled by their number). 
The sources with no detection of $v_7=1$ lines (condensations $9$, $10$, $11$ and $12$), only appear in the left-hand panels as lower limits for the $v_0/v_7$ ratio.

Table~\ref{tab:table_hcsumm} shows the estimated  parameters from the best fit to the non-LTE models  together with their associated errors. Given the number of free parameters in our non-LTTE modelling,  selecting the model parameters that best-fit  our data and estimating  the error is not straightforward. Fortunately, as illustrated in Fig.~\ref{fig:ratios},  the range of model parameters that can account for  the observed line ratios is relatively narrow.  Dust temperatures range from $200$ to $400$\,K,  in good agreement with  the LTE modelling, and densities are between $10^6$ and $10^7$\,cm$^{-3}$, with a systematic trend to lower densities when only the $v_0$\,$J=39-38/J=24-23$ ratio is considered.  The non-LTE parameters that \quotes{best fit}  the data  and their uncertainties have been derived from the parameter space of all models that fit the observed line intensity ratios within a given uncertainty of the observed ratio. For every SHC, we have extracted  the set of model parameters that fit, within $\pm20\%$,  the  observed $v_7$  line ratios. Then we have derived the best fit parameter  as the average of the  set of parameters weighted according to a Gaussian distribution. The errors shown in Table~\ref{tab:table_hcsumm} correspond to  the  sigma value of the weighted average. 

As previously mentioned, the inferred H$_2$ densities are very similar for all sources. However, the H$_2$ densities derived only from the $v_7 = 1$ rotational lines are systematically  larger than those from the ground state by a factor of $\sim 1-5$, indicating the presence of density gradients in the SHCs.  The non-LTE results from  Fig.~\ref{fig:ratios} shows that the H$_2$ densities are only weakly dependent on the kinetic  temperature,  $T_{\text{kin}}\sim T_{\text{dust}}$. Therefore, we have also derived the H$_2$ densities for sources $9$, $10$, $11$ and $12$, which are  also included Table~\ref{tab:table_hcsumm}, and  they are shown  Fig.~\ref{fig:ratios} by black arrows. The H$_2$ densities were derived from the measured line ratios of the rotational transitions in the ground vibrational state assuming a kinetic temperature close to the upper limit to T$_\text{vib}$ derived from LTE. It is remarkable that their H$_2$ densities are similar to those of the SHCs but with lower kinetic temperatures.

\begin{table*}
\begin{center}
\caption[]{\label{tab:table_hcsumm}Derived physical parameters for the SSCs in NGC\,253 from  LTE analysis and non-LTE modelling. Sizes are derived from $v_7=1$ ($v=0$ for sources in which $v_7=1$ transitions are upper limits, marked with $^*$) assuming it is optically thick and thus represent lower limits.
}
\begin{threeparttable}
\setlength{\tabcolsep}{4pt}
\begin{tabular}{ccccccccc}
\hline \noalign {\smallskip}
SSC	&   Type & 	Size	&$n_{H_2}$	 &$M_{H_2}$		  &T$_\text{vib}$  	&T$_\text{kin}$	&	L$_\text{app}\textsuperscript{a}$		&		L$_\text{app}\textsuperscript{b}$	\\
	&  	& (mas)	&  ($10^{6}$\,cm$^{-3}$)&  ($10^{3}$\,M$_\odot$)&  (K)	&	(K)	& 	($10^{8}$\,L$_\odot$) 		 &  ($10^{8}$\,L$_\odot$)	\\
\hline \noalign {\smallskip}
   	& 	&		&non-LTE		&non-LTE				&	LTE	   &	non-LTE	 & LTE &	non-LTE		\\ 
\hline \noalign {\smallskip}
1	    & SHC & $ 20$& {$1.9 \pm 0.4$}  & {$2.0 \pm 0.4$}	& $	216	\pm	15	$	& {$285 \pm 58$}   	&$	1.1 \pm 0.3  $	 	& 	{$	2.3  \pm 1.4$} \\
2	    & SHC & $ 22$& {$6.0 \pm 2.0$}  & {$8.0 \pm 2.7$}	& $	304	\pm	70	$	& {$322 \pm 64$}	&$	5.2 \pm 4.8  $ 		&   {$	4.4  \pm 2.5$} \\
3	    & SHC & $ 22$& {$1.6 \pm 0.6$}  & {$2.1 \pm 0.8$}	& $	337	\pm	50	$	& {$324 \pm 76$}	&$	7.9 \pm 4.6  $ 		&   {$	4.6  \pm 2.9$} \\
4	    & SHC & $ 19$& {$3.3 \pm 1.8$}  & {$3.0 \pm 1.7$}	& $	326	\pm	52	$	& {$191 \pm 34$}	&$	5.3 \pm 3.4  $ 		&	{$	1.1  \pm 0.4$} \\
5	    & SHC & $ 22$& {$4.5 \pm 1.9$}  & {$5.7 \pm 2.6$}	& $	269	\pm	22	$	& {$189 \pm 37$}	&$	3.1 \pm 1.0  $ 		&	{$	0.5  \pm 0.6$} \\
8	    & SHC & $ 25$& {$3.4 \pm 1.6$}  & {$6.8 \pm 3.2$}	& $	217	\pm	14	$	& {$263 \pm 38$}	&$	1.8 \pm 0.5 $		&	{$	3.5  \pm 1.4$} \\
9$^*$   &     & $ 26$& {$0.9 \pm 0.1$}  & {$0.6 \pm 0.1$}   & $ \leqslant 126$	& {$\leqslant 129$} &$	\leqslant 0.2 $    & $\leqslant 0.09$  \\
10$^*$	&     & $ 39$& {$10  \pm 3.3$}  & {$3.4 \pm 1.2$}   & $ \leqslant 132$	& {$\leqslant 118$} &$	\leqslant 0.6 $    & $\leqslant 0.06$  \\
11$^*$	&     & $ 36$& {$1.9 \pm 0.7$}  & {$2.1 \pm 0.8$}   & $ \leqslant 165$	& {$\leqslant 188$} &$	\leqslant 0.5 $    & $\leqslant 0.27$  \\
12$^*$	&     & $ 39$& {$1.0 \pm 0.4$}  & {$0.9 \pm 0.2$}   & $ \leqslant 140$	& {$\leqslant 125$} &$	\leqslant 0.6 $    & $\leqslant 0.13$  \\
13	    & SHC & $ 27$& {$1.6 \pm 0.2$}  & {$2.7 \pm 0.6$}	& $	393	\pm	34	$	& {$382 \pm 42$}	&$	23 \pm 8  $ 	& {$10 \pm 3$} \\
14	    & SHC & $ 42$& {$2.3 \pm 0.5$}  & {$22 \pm 4$}	& $	312	\pm	20	$	& {$271 \pm 47$}	&$	22 \pm 6  $ 	& {$10 \pm 4$} \\
\hline \noalign {\smallskip}
\end{tabular}
    \begin{tablenotes}
      \small
      \item \textsuperscript{a} Luminosities obtained assuming a black body emitting at the T$_\text{vib}$ derived from LTE modelling (i.e. $L=4\pi r^2 \sigma T_\text{vib}^4$).
      \item \textsuperscript{b} Luminosities obtained by integrating the predicted SED from the non-LTE models between $10$-$1200$\,$\mu$m.
    \end{tablenotes}
\end{threeparttable}
\end{center}
\end{table*}

\section{Derived properties}
\subsection{Sizes, HC$_3$N abundances and masses}
\label{subsec:sizes}

As already mentioned, since none of the detected SHCs are spatially resolved, we can set up an upper limit to their sizes of $0.1^{\prime \prime}$ ($<1.7$\,pc).
In addition, we can also estimate a lower limit to the SHCs sizes by assuming that the rotational transitions within the  $v_7=1$ state are optically thick and therefore the source brightness temperature will be the vibrational temperature derived from the LTE analysis. From the ratio of the observed and expected line intensities for the optically thick case and assuming a Gaussian source, we have estimated the lower limit to the sizes shown in Table~\ref{tab:table_hcsumm}. The lower limits range from $43$ to $14$ milliarcseconds (mas), which translate to $0.72$ and $0.24$\,pc. These lower limits are factors $2-7$ smaller than the upper limit to the size of $0.1^{\prime \prime}$.  

Combining the HC$_3$N column densities derived from the LTE analysis with the H$_2$ densities estimated from non-LTE modelling and the lower limits to their sizes, we can estimate the H$_2$ column densities, the fractional abundances of HC$_3$N, $X(\text{HC}_3\text{N})=N(\text{HC}_3\text{N})/N(H_2)$, and the masses of the forming SSCs. The estimated  H$_2$ columns densities range from $3-9 \times 10^{24}$\,cm$^{-2}$ and $X (\text{HC}_3\text{N})\sim 10^{-9}$, similar to the HCs found in our galaxy 
\citep[ 
$5\times 10^{-9}$ for Sgr B2M and Sgr B2N2, $5\times 10^{-9}$ for Orion KL HC, from ][]{deVicente2000, deVicente2002}.  From the H$_2$ column densities (N$_{\text{H}_2}$) and the H$_2$ densities derived from the non-LTE modelling we can estimate the depth of the emission along the line of sight. This depth, when compared to the estimated size, provides information on how the emitting regions are distributed along the line of sight. The mean value of the inferred depths is $0.89$\,pc, just within the lower and upper limits to the sizes, suggesting a nearly spherical distribution.

The H$_2$ masses of the SHCs in  Table~\ref{tab:table_hcsumm} range from a few $10^3$\,M$_\odot$ to a few $10^4$\,M$_\odot$. 
 However, the estimated masses from the models must be considered with caution since they are lower limits as they have been derived from the lower limit to the sizes. Nevertheless, the SHC masses derived from the HC$_3$N$^*$ emission only represent the hot inner core of the larger condensations  observed  by \citet{Leroy2018} in the $350$\,GHz continuum emission. Therefore, the M$_\text{gas}$ in Table~\ref{tab:table_comp_dust} is always larger than M$_{\text{H}_2}$ in Table~\ref{tab:table_hcsumm}, by up to a factor of $10$.

To derive the lower limit to the sizes of sources with no detection of HC$_3$N$^*$, we have used the same procedure as for SHCs, but using the line intensity of the rotational lines from the ground vibrational state and assuming a $T_{\text{kin}}$ of $130$\,K. The lower limits to the masses and the sizes are very similar to those derived for the SHCs.

\subsection{SSCs luminosities}
\subsubsection{\quotes{Apparent} luminosities}

From the estimated H$_2$ column densities from non-LTE modelling, the dust opacity in the mid-IR (the wavelength range responsible for the vibrational excitation of HC$_3$N) is larger than $20$ at $40$\,$\mu$m. 
Therefore, SHCs will emit as a black body at the temperature of the far-IR photosphere. A strong upper limit to the luminosity, which will be denoted as the apparent luminosity ($L_\text{app}$), can be obtained by adopting a temperature $T_\text{vib}$ for the photosphere, as derived from the LTE analysis. The same approach has been carried for the SSCs with only HC$_3$N$^*$ detection as upper limits. 
The derived $L_\text{app}$ from LTE modelling are shown in Table~\ref{tab:table_hcsumm}. 
In addition to the LTE estimates of the apparent luminosities, an alternative $L_\text{app}$ is also estimated from the integration between $10$ and $1200$\,$\mu$m of the non-LTE modelling predicted SED (Table~\ref{tab:table_hcsumm}). Both luminosities must be considered as lower limits since they were obtained by assuming the lower limit source sizes as derived from HC$_3$N$^*$ emission.
Similar apparent luminosities and trends are  found for both  LTE and non-LTE apparent luminosities. The difference between the apparent luminosities calculated from LTE and non-LTE, apart from the somewhat different vibrational/kinetic temperatures, arises from the fact that the LTE luminosity is from a black body and the non-LTE luminosity is calculated from a grey body, i.e. a factor $\sim 1.6$ between both luminosities.

\subsubsection{Protostar luminosities}
In the previous section we have made estimates of a lower limit to the SSCs apparent luminosities from the observed parameters. However, to estimate the actual luminosities of the heating sources is not straightforward. In fact, the estimated $L_\text{app}$ only represents the actual luminosity in the case of low dust column densities, i.e H$_2$ column densities of $< 10^{23}$\,cm$^{-2}$. For larger column densities, the derived $L_\text{app}$ should be considered an upper limit to the actual luminosity. This is due to the back-warming or greenhouse effect, first described by \citet[][]{Donnison1976} and more recently by \citet{GA19}.
The back-warming occurs when a fraction of the IR radiation from the heating source absorbed in an optically thick dust shell of the SHCs returns to the source due to the re-emission of the IR radiation by the inner surfaces of the shell. Then the thermal equilibrium at the inner surface is achieved for larger dust temperatures than those expected in the optically thin case. Therefore, the luminosity derived from the measured dust temperature at a given radius overestimates the actual luminosity of the heating source. For the large H$_2$ column densities found in the forming SSCs ($> 10^{24}$\,cm$^{-2}$) this effect needs to be taken into account to derive the luminosities of sources heating the SSCs.

\citet[][]{Ivezic1997} have made an estimate of the back-warming effect by using  self-similarity and the scaling method for a centrally heated spherical cloud for different density profiles. 
Following this method we can estimate the actual luminosity arising from the protostars in the SSCs ($\text{L}_{\text{p}^*}$)  can be inferred from $L_\text{app}$ by using equation $14$ of \citet{Ivezic1997} as:
\begin{align}
\label{eq:Lproto}
L_{\text{p}^*} = \frac{4}{\psi}L_\text{app}
\end{align}
where $\psi$ is  a complex function of the total column density and the radial density profile of the cloud, and of the emissivity properties of the dust. Since $\psi$ is a very strong function of the radial density profile, a precise estimate of the luminosity of the heating sources from the measured $L_{\text{app}}$ requires knowledge of the density gradient. 
Taking a representative H$_2$ column density of a few $10^{24}$\,cm$^{-2}$, as derived from the H$_2$ densities and the lower limit to the sizes, and considering that a HC typical density profile lies between $1$ and $2$, we estimate  the factor $4/\psi$ to be roughly of $1/10$. 
Consequently, in the following discussions we will consider the luminosity of the heating sources of the SSCs, $\text{L}_{\text{p}^*}$, to be around one order of magnitude smaller than the $\text{L}_{\text{app}}$ from Table~\ref{tab:table_hcsumm} ($\text{L}_{\text{app}}$ from the non-LTE modelling for SSCs with SHCs and from LTE modelling for the remaining SSCs). The rough estimated luminosities  $\text{L}_{\text{p}^*}$ associated to the protostars heating the SSCs are shown in Table~\ref{tab:table_comp_dust}. Taking into account that the luminosities have been derived from lower limit source sizes, they are likely to represent a lower limit to the actual luminosity. 
Considering the size derived from the depth of the emission  (see Section~\ref{subsec:sizes}), $\text{L}_{\text{app}}$ will be larger by a factor of $\sim 4$, but  the correction for the back-warming effect will also increase due to the larger H$_2$ column density. In this case the derived $\text{L}_{\text{p}^*}$ will be similar, within a factor of  $2$, to the $\text{L}_{\text{p}^*}$ derived from the lower limit to the sizes.

\begin{table*}
\begin{center}
\caption[]{\label{tab:table_comp_dust}SSCs sizes, masses and luminosities. Protostar luminosities (L$_{\text{p}^*}$) of condensations with no detection of $v_7=1$, marked with $^*$, have been calculated using T$_\text{rot}$ from the ground state rotational transitions and assuming a source size derived from its emission. Sources marked with $^{**}$ show no HC$_3$N emission. Masses in form of protostars (M$_{\text{p}^*}$) have been derived from L$_{\text{p}^*}$ assuming a light-to-mass ratio of $10^{3}$\,L$_\odot/$M$_\odot$, similar to the  value used for ZAMS massive stars.
}
\begin{threeparttable}
\setlength{\tabcolsep}{4pt}
\begin{tabular}{ccccccccccccc}
\hline \noalign {\smallskip}
 &	\multicolumn{2}{c}{Sizes}		&	&	 \multicolumn{3}{c}{Masses}		&	&	\multicolumn{3}{c}{Luminosities}	&		& SSC Phase  \\ \cmidrule{2-3} \cmidrule{5-7} \cmidrule{9-11} \cmidrule{13-13}
    
	&	SSC	&	Dust$_{350\text{GHz}}\textsuperscript{a}$&	& M$_\text{gas}$\textsuperscript{b}	&	 M$_{\text{p}^*}$	& M$_*$\textsuperscript{c}&		&	L$_{\text{p}^*}$	&	L$_*$\textsuperscript{d}	&	L$_{\text{p}^*}$/L$_*$&	&			\\
    &	(pc)	&	(pc) &	&		($10^4$\,M$_\odot$)	&	($10^5$\,M$_\odot$)	&	($10^5$\,M$_\odot$)&	& ($10^8$\,L$_\odot$)	&	($10^8$\,L$_\odot$)	&	&	&		\\
\hline \noalign {\smallskip}

1	&	0.34	&	2.7	&&	7.94	&0.3	&	0.20	&&	0.23	&	0.20	&	1.14	&&	proto		\\
2	&	0.37	&	1.2	&&	5.01	&0.4	&	0.20	&&	0.44	&	0.20	&	2.22	&&	proto		\\
3	&	0.37	&	2.6	&&	12.59	&0.4	&	0.13	&&	0.46	&	0.13	&	3.52	&&	proto		\\
4	&	0.33	&	2.5	&&	12.59	&0.1	&	1.00	&&	0.11	&	1.00	&	0.11	&&	proto		\\
5	&	0.37	&	2.1	&&	19.95	&0.05	&	2.51	&&	0.05	&	2.51	&	0.02	&&	ZAMS		\\
6	&	1.7$^{**}$&	2.1	&&	0.40	&	   	&	1.99	&&			&	1.99	&			&&	ZAMS		\\
7	&	1.7$^{**}$&	2.9	&&	3.16	&	   	&	0.32	&&			&	0.32	&			&&	ZAMS		\\
8	&	0.43	&	1.9	&&	15.85	&0.4	&	0.63	&&	0.35	&	0.63	&	0.56	&&	proto		\\
9$^*$	&	0.44&	2.6	&&	5.01	&0.02   &	3.16	&&  0.02  	&	3.16	&	$<0.01$	&&	ZAMS		\\
10$^*$	&	0.67&	3.5	&&	15.85	&0.06   &	1.99	&&  0.06    &	2.00	&	$<0.03$	&&	ZAMS		\\
11$^*$	&	0.62&	2.9	&&	3.16	&0.05   &	3.98	&&  0.05    &	3.98	&	$<0.01$	&&	ZAMS		\\
12$^*$	&	0.67&	4.3	&&	1.26	&0.06   &	10.00	&&  0.06    &	10.00	&	$<0.01$ &&	ZAMS		\\
13	&	0.46	&	1.6	&&	15.85	&1.0	&	0.63	&&	1.02	&	0.63	&	1.62    &&	proto		\\
14	&	0.72	&	1.6	&&	50.12	&1.0	&	3.16	&&	1.00	&	3.16	&	0.32 	&&	proto		\\
\hline \noalign {\smallskip}
\end{tabular}
\begin{tablenotes}
      \small
      \item \textsuperscript{a}\citet[][]{Leroy2018} sizes derived from the dust continuum emission at $350$\,GHz.
      \item \textsuperscript{b}\citet[][]{Leroy2018} gas mass estimates from  $350$\,GHz dust emission, assuming T$_\text{dust}=130$\,K and a dust-to-gas ratio of $0.01$.
      \item \textsuperscript{c}\citet[][]{Leroy2018} ZAMS stellar masses derived from L$_\text{p*}$ assuming a light-to-mass ratio of $10^3$L$_\odot/$M$_\odot$.
      \item \textsuperscript{d}\citet[][]{Leroy2018} derived luminosities from the $36$\,GHz continuum emission assuming it is dominated by free-free emission.
    \end{tablenotes}
\end{threeparttable}
\end{center}  
\end{table*}

\section{Discussion}
\label{sec:Discussion}

\subsection{SHCs in NGC\,253: Evolutionary earliest phases of Super Star Clusters}

SSCs represent the most massive ($\text{M}_* \sim 10^5 -10^7$\,M$_\odot$) examples of clustered star formation. SSCs are compact, with radius $\sim 1-5$\,pc, and young, $1-100$\,Myr \citep[][]{Whitmore2002, Alonso-Herrero2003, Kornei2009}. Although SSCs are believed to be the precursors of Globular Clusters (GCs, with ages $\gtrsim 10$\,Gyr), not all SSCs will evolve to a bound cluster 
as it requires a very high star formation efficiency (SFE) and high star formation rate (SFR) to prevent early disruption of the cocoon due to the feedback generated by high mass star formation \citep[][]{Hills1980, Beck2015, Johnson2015}.

HCs are indeed expected to represent the earliest phases of massive star formation.
HCs are internally heated by massive protostars deeply embedded in the parent molecular cloud, which is still undergoing gravitational collapse with mass accretion rates as high as a few $10^{-3}$M$_\odot$\,yr$^{-1}$  \citep[][]{Walmsley1995}. The consequent large concentration of gas and dust around protostars prevents the development of the Ultra-Compact \ion{H}{ii} region (UCHII) \citep[][]{Walmsley1995, Churchwell2002, Hoare2007}.
The  UCHII emerges when the accretion rate decreases below a threshold value. The flickering of the continuum emission observed in UCHII has been interpreted as due to the latest episodes of mass accretion onto UCHII regions \citep[][]{dePree2014}. Once the accretion stops, the UCHII region expands and evolves into an \ion{H}{ii} region.
The timescales for these processes are very short, from $\sim 10^5$\,yr when the  \ion{H}{ii} region may start to show up, to a few $10^6$\,yr, when SN explosions from the most massive stars will take place. 

Our detection of HC$_3$N$^*$ emission, indicative of extremely high column densities of gas and dust around the heating protostars, suggest that the condensations detected in the continuum by \citet[][]{Leroy2018} represent indeed the earliest phases of SSCs evolution.
In fact, some of them ($10$, $11$ and $13$) show strong Hydrogen recombination lines likely associated to \ion{H}{ii} regions with a steep electron density profile  \citep[][]{Baez2018}, as expected for extremely young UCHII regions \citep[][]{Jaffe1999, Baez2014}. 
We have detected $8$ out of $14$ SSCs candidates in HC$_3$N$^*$ (i.e. SHCs), indicative of a very early phase in their evolution.

Table~\ref{tab:table_comp_dust} summarizes the main properties of the NGC\,253 forming SSCs, their sizes, their stellar and gas content and their luminosities.
 The gas mass (M$_\text{gas}$) of the SSCs, derived from the $350$\,GHz dust continuum emission; the luminosities (L$_*$) from ionizing Zero Age Main Sequence (ZAMS) stars, estimated from the $36$\,GHz continuum emission assuming it is dominated by free-free emission; and the mass of ZAMS stars (M$_*$), obtained from L$_*$ by assuming a light-to-mass ratio of $10^3$\,L$_\odot/$M$_\odot$; have been taken from \citet{Leroy2018}.
The L$_*$ and M$_*$ values could be overestimated in some sources in the very central region 
 \citep[close to the brightest radio source TH2,][]{Turner1985}  due to  the contribution of synchroton emission to the $36$\,GHz continuum emission \citep{Baez2018}.
 To complete the census of the star population in the forming SSCs, we have also added the luminosity in protostars, L$_\text{p*}$ (i.e. the apparent luminosities from Table~\ref{tab:table_hcsumm} corrected by an order of magnitude using Eq.~\ref{eq:Lproto}). 
The estimated protostellar luminosities are typically of a few $10^7$\,$\text{L}_\odot$ for most of the SHCs, with SHC13 and SHC14 reaching  $10^8$\,$\text{L}_\odot$. On the other hand, sources with HC$_3$N$^*$ only detected as upper limits have estimated protostellar luminosities $\sim10^6$\,$\text{L}_\odot$, one order of magnitude smaller than the SHCs. 
The total luminosity (from proto and ZAMS stars) of all SSCs  is  $3.4\times 10^9$\,$\text{L}_\odot$ (from which L$_\text{p*Total}=4\times10^8$\,$\text{L}_\odot$ and L$_\text{*Total}=30\times10^8$),  which accounts for about $22\%$ ($\sim3\%$ from protostars and $\sim19\%$ from ZAMS stars) of the total luminosity of the central region of NGC\,253, assuming that half of the galaxy's total luminosity, $3.1 \times 10^{10}$\,$\text{L}_\odot$, arises form the central $170$\,pc \citep[][]{Melo2002,GA15}.
The remaining central luminosity of the galaxy would be produced by the star formation and more evolved SSCs outside the studied condensations (see Fig.~\ref{fig:SSC_pos}) that occurred in the last $10$\,Myr \citep{Watson1996, Fernandez-Ont2009}. 

We have used the luminosity to make an estimate of the mass in  protostars, M$_\text{p*}$, by assuming a light-to-mass ratio of $10^3$\,L$_\odot$\,M$_\odot^{-1}$. This is the same value used by  \citet{Leroy2018} to derive the mass in ZAMS stars (M$_\text{*}$). 
We used this value since the timescales for the massive protostars to reach the ZAMS are rather short and are expected  to be close to the ZAMS evolutionary track \citep{Hosokawa2009}. The assumed light-to-mass ratio  is obviously uncertain since we do not know the SSC Initial Mass Function and the accretion rates. We have adopted the value corresponding to a representative cluster star of $10-20$\,M$_\odot$ with a high accretion rate of a few $10^{-4}$\,M$_\odot$\,yr$^{-1}$. Our adopted light-to-mass ratio is also close to that required for radiation pressure support \citep{GA19} and to that of $2.1\times10^3$\,L$_\odot$\,M$_\odot^{-1}$ obtained for the 30 Doradus region by \citet{Doran2013}. Most of the following discussions will not be affected by this assumption since it will affect all SSCs in a similar  way.
 
We have classified the SSCs into proto-dominated SSCs (hereafter, proto-SSCs) and ZAMS-dominated SSCs (ZAMS-SSCs) by comparing the luminosities arising from proto and ZAMS stars for each SSCs. SSCs with L$_\text{p*}$/L$_\text{*}>0.1$ are classified as proto-SSCs (which are the same SSCs containing SHCs except for source 5) and those with L$_\text{p*}$/L$_\text{*}<0.1$ as ZAMS-SSCs.

 The mass in protostars, M$_\text{p*}$ ranges from $0.1-1.0 \times 10^5$\,M$_\odot$ for proto-SSCs and from $2-6 \times 10^3$\,M$_\odot$ for ZAMS-SSCs. This is in contrast with the mass in ZAMS stars in the SSC candidates, which seems to be anti-correlated with the mass in the protostar phase. Figure~\ref{fig:distances} shows in the middle panel the  L$_\text{p*}$/L$_\text{*}$ ratio (i.e. M$_\text{p*}$/M$_\text{*}$) as a function of the distance of the forming SSCs to TH2  \citep[][]{Turner1985}, which is illustrated in the upper panel of the figure. For sources $6$ and $7$ we have adopted a fiducial value of L$_\text{p*}$/L$_\text{*}=0.001$ since no estimation of L$_\text{p*}$ was made.

\subsubsection{Age of SSCs} 
\begin{figure}
\centering
    \includegraphics[width=\linewidth]{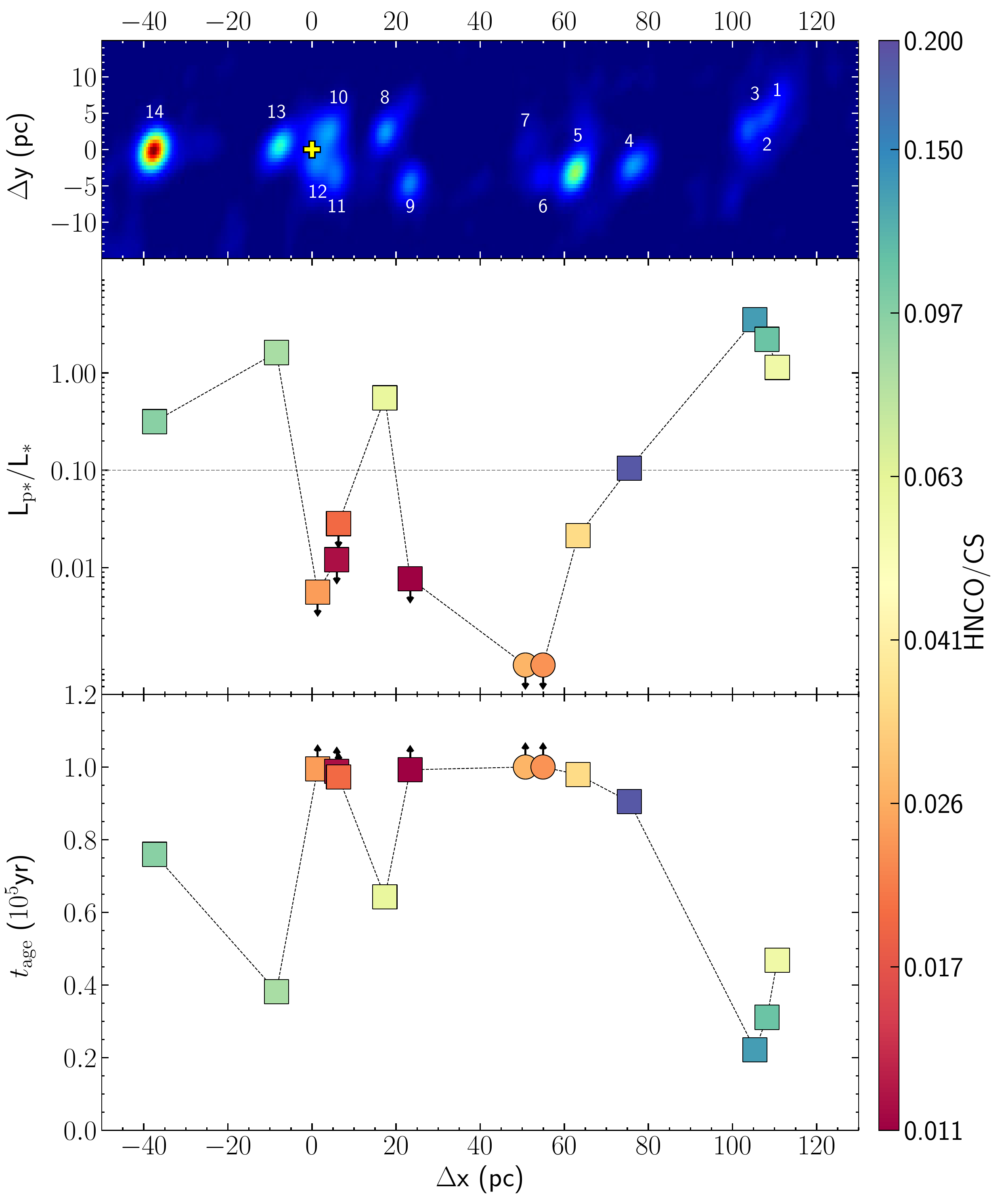}
      \caption{
      The top panel shows the rotated NGC\,253 $218$\,GHz continuum emission. TH2 is marked with a yellow cross.
      The middle panel shows proto and ZAMS stars luminosity ratio against the distance of each condensation to TH2. ZAMS-SSCs are indicated as upper limits. For sources $6$ and $7$, represented with circle, we have adopted a fiducial value of $\text{L}_\text{p*}/\text{L}_{*}=0.001$. The lower panel shows a rough estimation of the SSCs age based on its luminosity ratio $\text{L}_\text{p*}/\text{L}_{*}$ and assuming the timescale of an \ion{H}{II} region to be $10^5$\,yr. Sources are colored by their HNCO/CS ratio.}
      \label{fig:distances}
\end{figure}

The L$_\text{p*}$/L$_\text{*}$ ratio  in Figure~\ref{fig:distances} shows a clear trend, as it varies  from  $0.1-3.5$ for the proto-SSCs to  $0.005-0.01$ for those that are ZAMS-SSCs. The large changes in this ratio, by up to $3$ orders of magnitude, can be related to the evolutionary stage of the SSCs. In the current picture of massive star formation, the SHC phase indicates that star formation is still going on, and the lack of SHCs associated with very young UCHII regions suggests that mass accretion has ended and the SSCs are completing their formation. Then, it is expected that the L$_\text{p*}$/L$_\text{*}$ ratio should be roughly related to the age of the SSCs. It is accepted that SSCs are very short-lived ($\lesssim 10^{5 -6}$\,yr)  before they  start to disrupt their natal molecular cloud and show up in the visible range \citep[][]{Johnson2015}.
Then, considering that the time scale for UCHII regions to become optically thin is a few $10^5$\,yr we will assume that SSCs will be completely formed in about $\sim 10^5$\,yr.  Under this assumption, a rough estimate of the SSCs age can be obtained as follows:
\begin{align}
\label{eq:tage}
t_\text{age}(\text{yr}) =
\begin{cases}
\frac{1}{1+\text{L}_{\text{p*}}/\text{L}_*} \times 10^5 &,\, \text{for}\,\,  \text{L}_{\text{p*}}/\text{L}_* \geqslant 0.05\\
\\
\gtrsim 10^5 &,\, \text{for}\,\, \text{L}_{\text{p*}}/\text{L}_* < 0.05\\
\end{cases}
\end{align}

which is shown in the lower panel of Fig.~\ref{fig:distances} for all the SSCs. The estimated age of the SSCs will scale with the assumed timescale of their formation. We find that SHC$3$ is the youngest proto-SSC, while source $5$ (detected in HC$_3$N$^*$) would be already in the ZAMS phase. 
The short timescales of the HCs ($\lesssim 10^5$\,yr) could be an explanation of the lack of widespread detection of HCs in galaxies so far \citep{Martin2011, Shimonishi2016, Ando2017}. 
 Again, we have to treat sources $11$ and $12$ with caution, because they are in a complex region where  a significant contribution from non-thermal emission may be present and hence their $\text{L}_{*}$ and $\text{M}_{*}$  could have been overestimated, as indicated by \citet{Leroy2018}.

The estimated age of the SSCs ($t_\text{age}$) in Fig.~\ref{fig:distances} shows a clear trend in their evolutionary stage as a function of their position. 
Central condensations in  Fig.~\ref{fig:distances} ($4$, $5$, $6$, $7$, $9$, $10$, $11$ and $12$) are more evolved than the sources at the edges ($1$, $2$, $3$, $13$ and $14$). The exception of source $8$ could be explained if it were only apparently close to the center due to a projection effect. 

\subsubsection{Radiative feedback}

 Massive stars have a strong impact on their surroundings due to both mechanical and radiative feedback. Once the massive stars in the SSCs reach the UCHII region phase it is expected that the UV radiation will affect the surrounding material creating photodissociation regions (PDRs).  Then one expects that the difference in  the evolutionary stage found  in the SSCs in NGC\,253 will have an impact in the chemical richness of the molecular gas in SSCs. In fact,  \citet{Ando2017} found that at scales of $10$\,pc (at a lower resolution than in this work)  the  HNCO and CH$_3$OH abundances dramatically decreases in two of their sources, which actually contain our ZAMS-SSCs $10$, $11$, $12$ and $9$. \citet{Martin2008, Martin2009} have found that the HNCO/CS ratio is an excellent  tracer of gas affected by UV radiation since HNCO is much more easily dissociated than CS (which is still abundant in PDRs). The rather low relative abundances of HNCO at scales of $10$\,pc suggests that in ZAMS-SSCs  the radiation from the newly formed O-type stars have already created PDRs, destroying a large fraction of molecular gas in their surroundings.
  
\begin{figure*}
\centering
    \includegraphics[width=\linewidth]{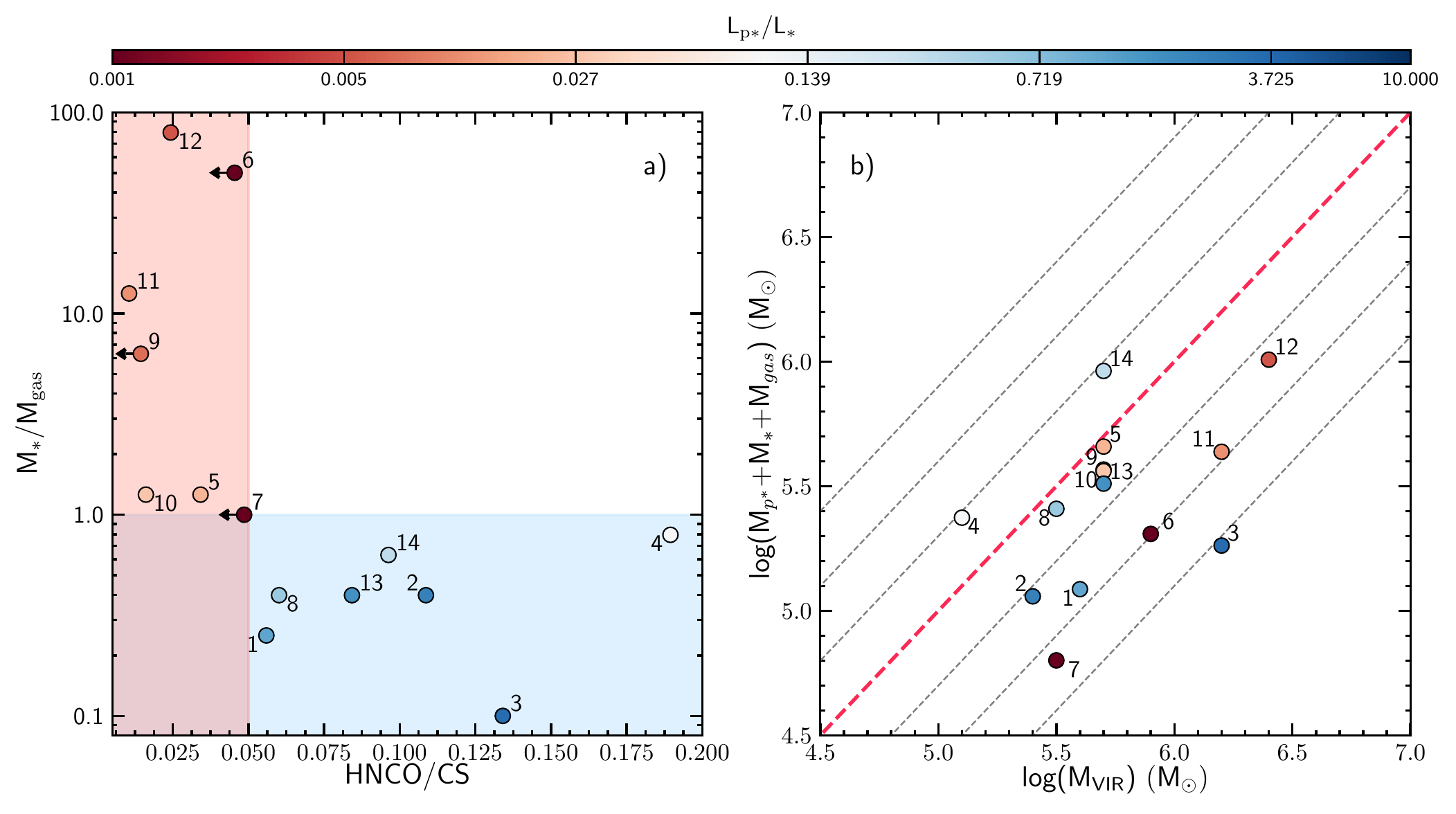}
      \caption{a) Radiative feedback: ratio between mass in form of stars and gas mass against the integrated intensity HNCO/CS ratio (i.e. radiative feedback). Blue and red shaded regions contains SSCs with $\text{L}_{\text{p*}}/\text{L}_*>0.1$ and $<0.1$, respectively.  b) Mechanical feedback: virial mass  from \citet{Leroy2018} against total mass (mass in gas, stars and protostars) for each SSC candidate. Sources are colored by their $\text{L}_{\text{p*}}/\text{L}_*$ ratio.
      }
      \label{fig:subfig_ratios_feedback}
\end{figure*}

In order to better quantify the effect of the radiative feedback in the SSCs, we have derived from our data set the HNCO/CS ratio with the same angular resolution as the HC$_3$N data by using  the integrated intensities of the HNCO\,$(16_{1,15}-15_{1,14})$ and CS\,$(7-6)$ lines towards all the SSCs, see Table~\ref{tab:hnco_cs}. 
 Panel a) of Figure~\ref{fig:subfig_ratios_feedback} displays this ratio against the ratio between the stellar mass and the gas mass, $\text{M}_{*}/\text{M}_\text{gas}$. The HNCO/CS ratio is expected to be inversely related to the ratio of the number of UV photons (stellar mass) and the total gas mass.
 
 \begin{table}
\begin{center}
\caption[]{\label{tab:hnco_cs}Observed integrated line emission in Jy\,beam$^{-1}$\,km\,s$^{-1}$ for HNCO and CS lines.}
\setlength{\tabcolsep}{6pt}
\begin{tabular}{ccccc}
\hline \noalign {\smallskip}
HC	&	 FWHM HNCO	&	HNCO	& FWHM CS &	CS		\\
	& $(\rfrac{\text{km}}{\text{s}})$ &\scriptsize	$16_{1,15}-15_{1,14}$& $(\rfrac{\text{km}}{\text{s}})$	& \scriptsize	$7-6$	\\ 
\hline \noalign {\smallskip}
 \scriptsize $\nu$\,(GHz)	& 	&\scriptsize 352.90	&	& \scriptsize	342.88		\\
\scriptsize $E_{LO}$\,(K) &	& \scriptsize	118.37	& 	& \scriptsize	34.32	\\
\hline \noalign {\smallskip}
1	& $	17.0 $ & $	0.117	        $ & $	42.0	$ & $	2.09	$ \\
2	& $	33.2 $ & $	0.275	        $ & $	49.6	$ & $	2.53	$ \\
3	& $	23.0 $ & $	0.272	        $ & $	57.3	$ & $	2.03	$ \\
4	& $	27.8 $ & $	0.226	        $ & $	36.2	$ & $	1.44	$ \\
5	& $	44.0 $ & $	0.267	        $ & $	41.4	$ & $	7.81	$ \\
6	& $	38.0 $ & $ \leqslant 0.046	$ & $	38.0	$ & $	1.01	$ \\
7	& $	40.0 $ & $ \leqslant 0.043	$ & $	39.6^*	$ & $	0.88	$ \\
8	& $	24.5 $ & $	0.160	        $ & $	37.0	$ & $	2.66	$ \\
9	& $	33.2 $ & $ \leqslant 0.060 	$ & $	37.6^*	$ & $	2.08	$ \\
10	& $	34.8 $ & $	0.055	        $ & $	35.5^*	$ & $	2.84	$ \\
11	& $	50.8 $ & $	0.065	        $ & $	56.4^*	$ & $	5.45	$ \\
12	& $	40.0 $ & $	0.082	        $ & $	55.8^*	$ & $	3.37	$ \\
13	& $	32.0 $ & $	0.293	        $ & $	46.0^*	$ & $	3.48	$ \\
14	& $	45.0 $ & $	1.463	        $ & $	42.0^*	$ & $	15.19	$ \\
\hline \noalign {\smallskip}
\end{tabular}
\begin{tablenotes}
      \small
      \item $^*$ Two CS components or CS autoabsorption.
    \end{tablenotes}
\end{center}
\end{table}

Figure~\ref{fig:subfig_ratios_feedback}a clearly shows two different regimes shown as blue ($\text{L}_{\text{p*}}/\text{L}_*$>0.1) and red ($\text{L}_{\text{p*}}/\text{L}_*$<0.1)  shaded regions. The SSCs located in the red region, with large masses in stars as compared to their gas mass
($\text{M}_{*}/\text{M}_\text{gas} \gtrsim 1$), have already photo-dissociated most of their HNCO. Sources with HNCO/CS values $\lesssim 0.05$ ($12$, $11$, $10$, $9$, $7$, $6$ and $5$) are indeed the oldest and more evolved ones, with estimated ages around $\gtrsim 10^5$\,yr. 
This result strongly suggests  that radiative feedback has started to have an important effect in the chemical properties of the molecular gas left after star formation and it might have played a role in quenching the star formation on these sources. This is in agreement with their rather low $\text{L}_{\text{p*}}/\text{L}_*$ ratio of $<0.1$. On the other hand, the  SSCs  in the blue region with higher $\text{L}_{\text{p*}}/\text{L}_*$ and low $\text{M}_{*}/\text{M}_\text{gas}$ show rather high HNCO/CS, as expected if the UV radiation from the (few) recently formed stars do not significantly affect the chemical properties of the molecular gas due to the relatively low amount of stars in the ZAMS phase as compared to that in the protostar phase (high $\text{L}_{\text{p*}}/\text{L}_*$  ratios) or because it is very well shielded. 
This is expected for very young SSCs still forming stars, where the PDRs created by a still low fraction of massive stars in the ZAMS phase represents a small fraction of the total gas mass. Within this context, source $4$ would be in an intermediate state between these two phases, with high HNCO/CS ratio but rather low $\text{L}_{\text{p*}}/\text{L}_*$; and source $5$ would have reached the ZAMS phase recently.
In summary, the overall chemical effects revealed by the HNCO/CS ratios suggest that it is consistent with the picture of the formation and evolution of SSCs, indicating that the  radiative feedback effects appear relatively quick in just a few $10^5$\, yr, when the SSCs seem to be completely formed. 

\subsubsection{Mechanical Feedback}

So far it is unclear whether the mechanical feedback produced by the massive stars in the proto and ZAMS phases plays a significant role in the very early stages of SSCs evolution. While radiation could  permeate the whole SSC molecular cloud in very short timescales, depending on the number of ionizing stars and the extinction (i.e. the total dust column gas), the mechanical feedback is expected to have a longer time scale to have a sizeable effect on the whole cloud. This is due to the nature of mechanical feedback which is injected in the cloud by the winds of the proto and ZAMS stars in the SSCs. Usually, mechanical feedback is observed by means of P-Cygni profiles \citep{GA2012} or broad  wings in the molecular line profiles.
Alternatively, we can use the Virial Theorem to look how the total kinetic energy relates to the total potential energy of the SSCs. \citet{Leroy2018} have already discussed the dynamical mass and the stellar and gas content of the SSCs. We have updated the \citet{Leroy2018} Virial analysis by adding the protostar component found in this work to their stellar mass as shown in panel b) of Fig.~\ref{fig:subfig_ratios_feedback}. Some of the proto-SSCs are close to virialization ($5$, $8$, $13$), but most of them are not virialized.
Although older ZAMS-SSCs seem to have a larger non-virialized state ($6$, $7$, $9$, $11$, $12$), as expected from mechanical feedback, also very young proto-SSCs (like $1$ and $3$) have  larger dynamical masses than that in gas and stars. Hence, even taking into account the protostar component, the Virial analysis does not show any clear trend on the dynamical state of the SSCs related to mechanical feedback as one would expect from their evolution.

\subsection{Star formation efficiency in the SSCs}
 
\begin{figure*}
\centering
    \includegraphics[width=\linewidth]{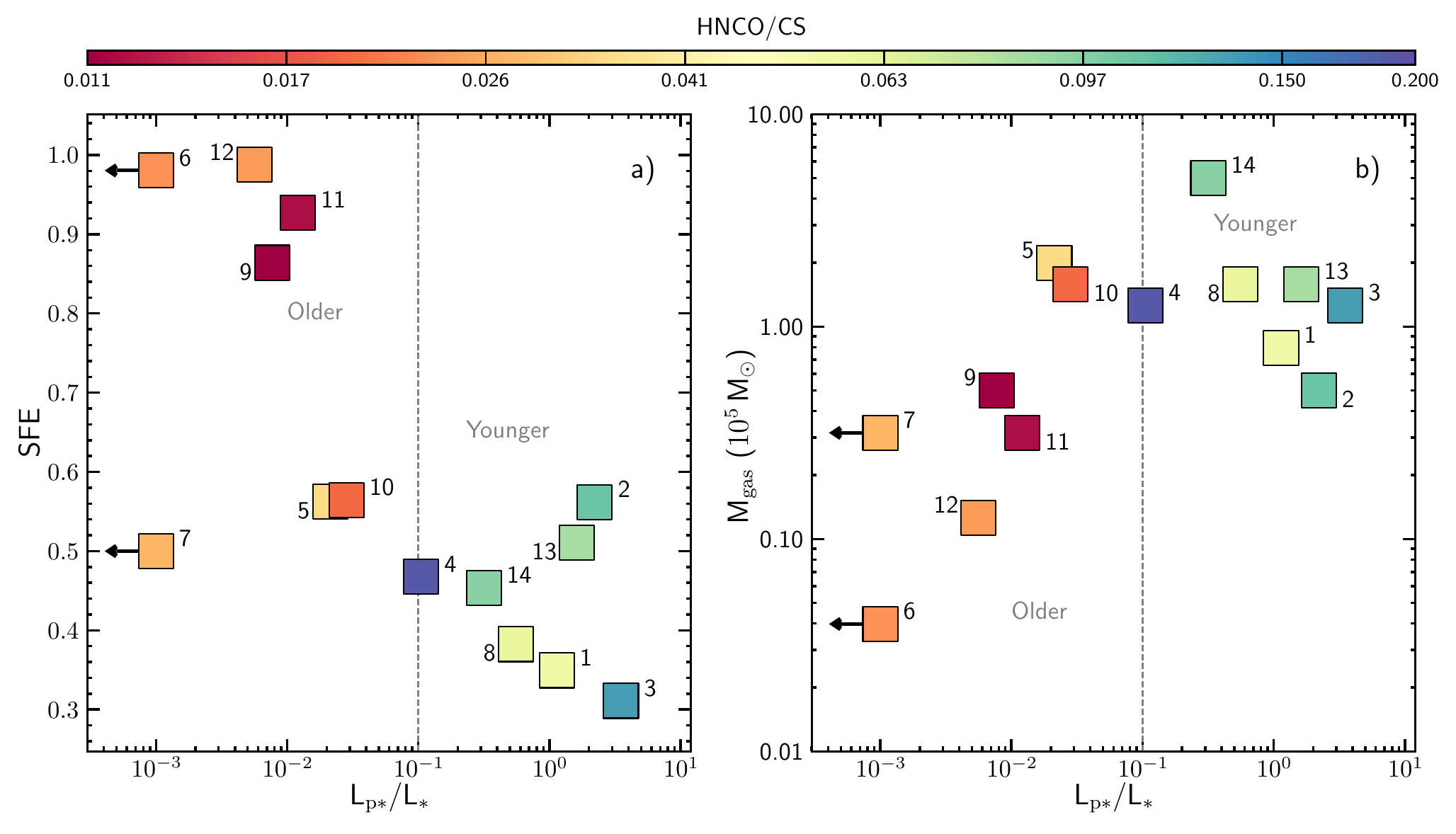}
      \caption{a) Star Formation Efficiency against luminosity ratio. b) SSC candidates gas mass versus their luminosity ratio. Sources are colored by their HNCO/CS ratio. The luminosity and the HNCO/CS ratios are used as proxies of the SSCs age.
      }
      \label{fig:subfig_ratios_SFE}
\end{figure*}

The SFE of a molecular cloud with an initial mass $\text{M}_\text{initial}$ is given  by the ratio between the mass converted into stars and the initial mass,  ($\text{M}_{p*}+\text{M}_{*})/\text{M}_\text{initial}$. Assuming that there has not been significant mass loss, as inferred from the discussion above on the radiative and mechanical feedback, $\text{M}_\text{initial}$ can be estimated from the sum of the mass already in  stars plus the remaining gas mass ($\text{M}_{p*}+\text{M}_{*}+\text{M}_\text{gas}$),  will be given by 
\begin{align}
\frac{1}{1+\text{M}_\text{gas}/(\text{M}_{p*}+\text{M}_{*})}
\end{align}
In case that the mechanical feedback is not negligible, the derived SFEs must be considered as upper limits. Using this approach, we have estimated the SFEs shown in panel a) of  Fig.~\ref{fig:subfig_ratios_SFE} as a function of the age of the SSCs (derived from their L$_{p*}$/L$_*$ ratio), colored by their HNCO/CS ratio. It is remarkable that the SSCs show two different SFE regimes. The  young  proto-SSCs, including intermediate sources like $5$ and $10$ with  SFEs of  $\sim 30-60\%$ and the more evolved ZAMS-SSCs with SFEs   $>85\%$. SSC $7$ is clearly outside this trend. It could be that it was not massive enough  to maintain a high SFE, it has the lowest $\text{M}_\text{initial}=6.3\times 10^4$\,M$_\odot$ while the other SSCs have  $\text{M}_\text{initial}>1.2\times 10^5$\,M$_\odot$.

The higher SFEs derived for the ZAMS-SSCs are in accordance with their evolutionary stage, since they are more evolved and have had more time to convert gas into stars.
This is further supported by the correlation we have found (see  Fig.~\ref{fig:subfig_ratios_SFE}b) between gas mass and age (i.e.  L$_{p*}$/L$_*$) and also the radiative feedback (HNCO/CS ratio). 
In addition, if the  proto-SSCs continue transforming gas into stars will finally also achieve a very high SFE of  $>85\%$. 
A high SFE means that  most of the stars have to be formed in a very short time scale because the feedback (radiative or mechanical) generated by the stars soon starts to halt the star formation. In addition, as discussed below, the \quotes{future} SFR should be high enough to complete the conversion of a large fraction of gas mass into stars (with the exception of source $7$).

\subsection{Evolution of the star formation rates during the SSC formation}

\begin{figure}
\centering
    \includegraphics[width=\linewidth]{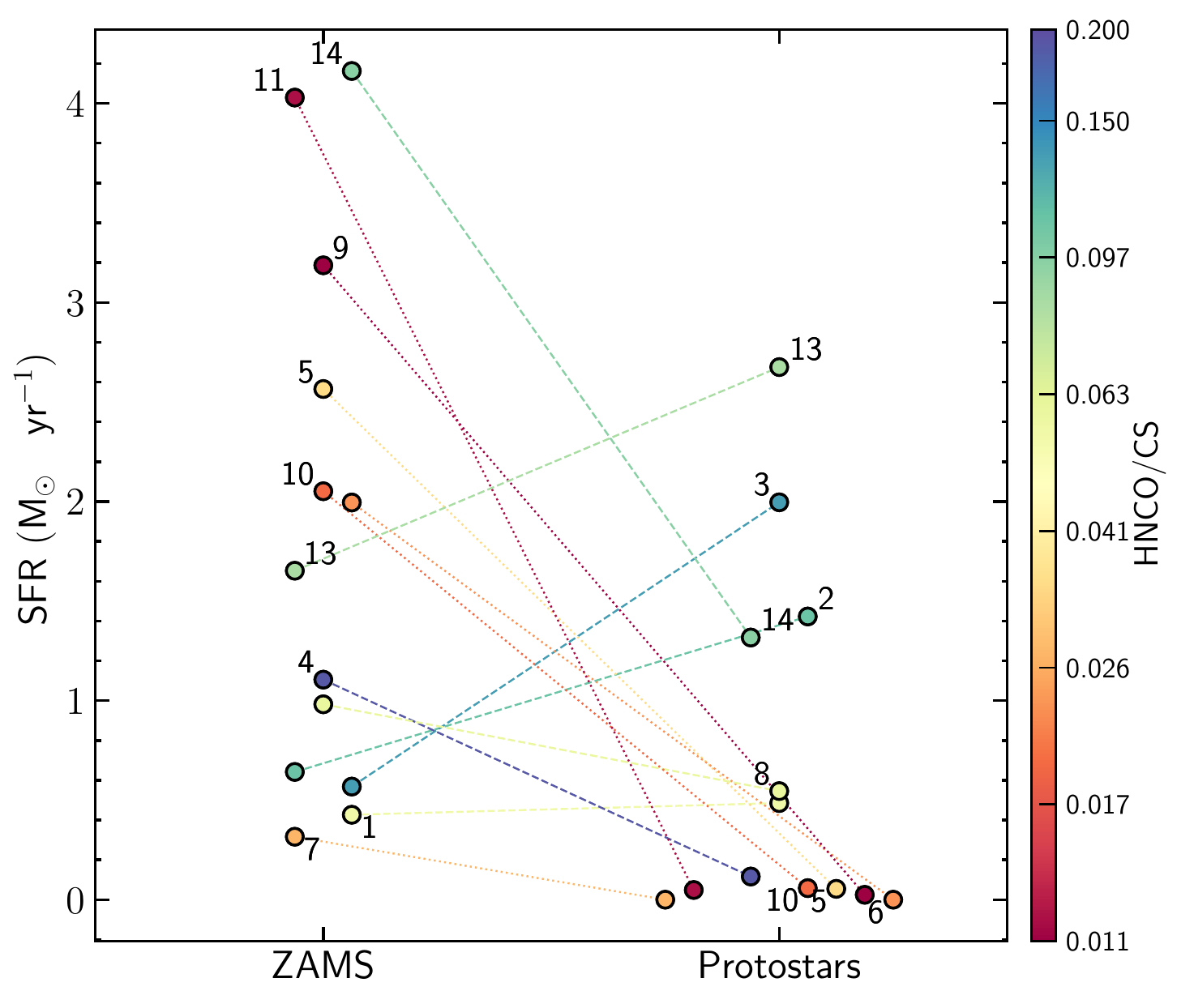}
      \caption{Derived SFRs for the different phases of SSCs. The ZAMS and protostars SFRs have been derived from the masses of the two stellar components  and their estimated ages ($\text{M}_{*}/t_\text{age}$ and $\text{M}_{p*}/t_\text{age}$, respectively) .
      }
      \label{fig:SFR}
\end{figure}

From the lifetimes and the stellar masses involved in the different phases of the formation of the SSCs, we can estimate the history of the SFRs during their formation.  
Let us first consider the SFRs required to form the stellar components in 
the proto and ZAMS star phases for all SSCs. 
Figure~\ref{fig:SFR} shows the estimated SFRs   for the stellar components as traced by the protostars ($\text{M}_{p*}/t_\text{age}$) and  by the ZAMS stars ($\text{M}_{*}/t_\text{age}$), colored by their HNCO/CS ratio. 
The SFRs derived from ZAMS stars span from $\sim 0.5$ to $5.5$\,M$_\odot$\,yr$^{-1}$, not showing any systematic trend. For instance, SSCs $11$ and $14$,
ZAMS and proto-SSCs respectively, show similar high ZAMS SFRs of $4$ and $5.5$\,M$_\odot$\,yr$^{-1}$.
The SFRs derived for the protostars only applies to proto-SSCs, which are still forming stars and range from $\sim 2$ to $6$\,M$_\odot$\,yr$^{-1}$. The protostar SFRs of the ZAMS-SSCs is close to $0$. Most of the proto-SSCs ($1$, $2$, $3$, $13$) show somewhat higher protostar SFRs than ZAMS SFRs. However, sources $14$, $8$ and $4$ show just the opposite behaviour. However, given the uncertainties in our estimates we consider that the SFRs did not change between both phases. We can make a projection of the expected SFRs by considering that the final SFE of the proto-SSCs will be of about $0.95$, similar to that of the ZAMS-SSCs. 
It is noteworthy that basically all proto-SSCs require to maintain similar SFRs than those found in the previous phases to achieve a SFE of  $\sim 0.95$. The only exception would be SSC $14$, which would require a very high SFR of $\sim 4$\,M$_\odot$\,yr$^{-1}$  to form the first massive stars  already in the ZAMS phase; then decrease to $1$\,M$_\odot$\,yr$^{-1}$ for the newly formed stars (protostars in the SHCs); and finally would have to increase its SFR up to $10$\,M$_\odot$\,yr$^{-1}$ to archive a SFE of $\sim 0.95$ in $10^5$\,yr. 

The SFRs estimated for the different phases seem to be independent from the SSCs age or evolutionary phase. The SFRs of the SSCs ranges from $1$ to $4$\,M$_\odot$\,yr$^{-1}$,  with the exception of SSCs $11$ and $14$, which show a SFR of $>4 $\,M$_\odot$\,yr$^{-1}$. These are the only SSCs that show a SFR higher than the global value of $3-4$\,M$_\odot$\,yr$^{-1}$ \citep{Ott2005, Bendo2015}.
While most of the proto-SSCs likely  achieve SFEs of $\sim 1$, the evolution of the SSC $14$ is less clear since it will require a substantial increase in the SFR in the next few $10^4$\,yr, however the radiative feedback is still negligible (high HNCO/CS ratio) and suggest that star formation could not be quenched in less than $10^4$\,yr.

\subsection{On the evolution of SSCs in galaxies}

\begin{figure}
\centering
    \includegraphics[width=\linewidth]{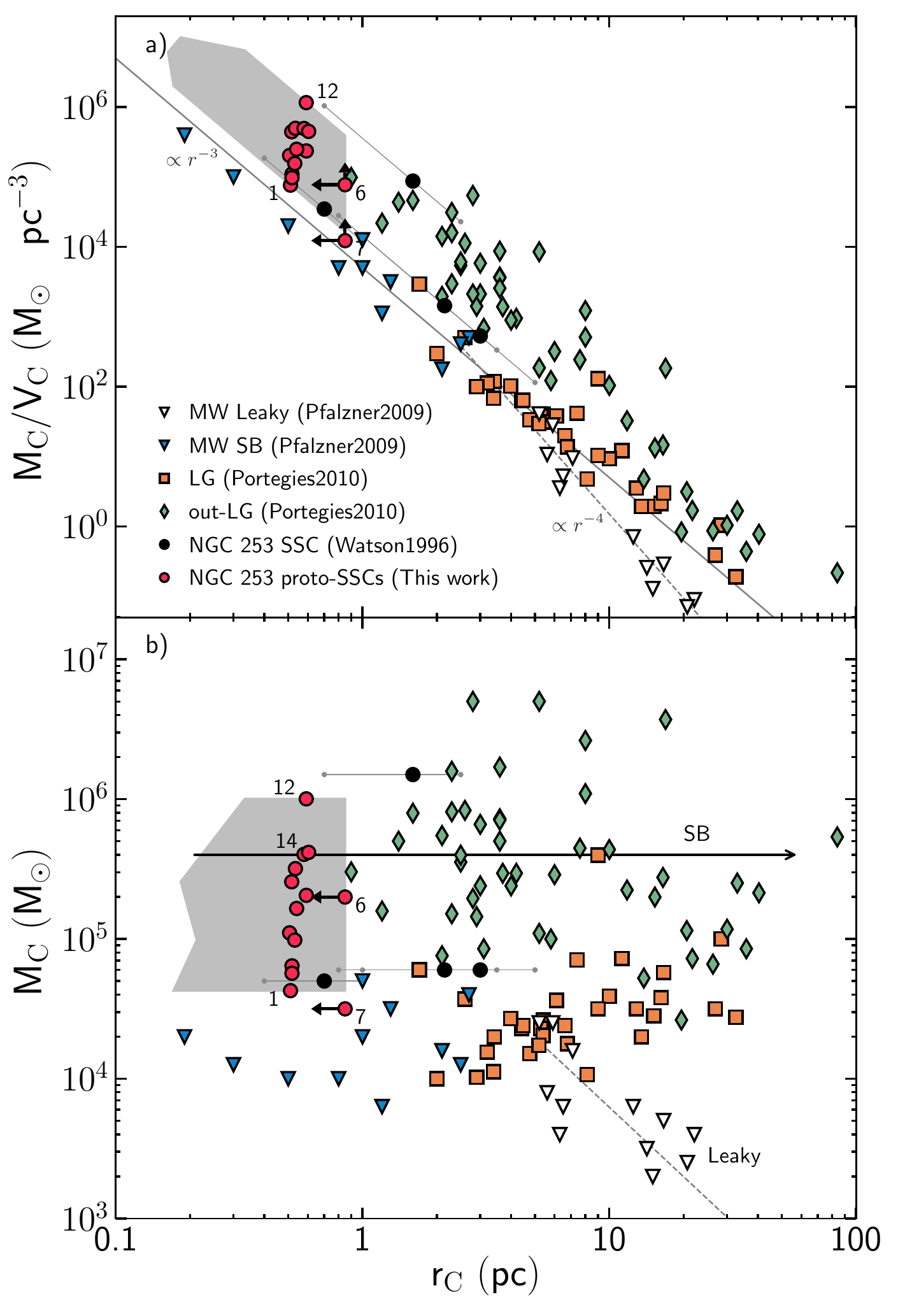}
      \caption{Comparison of the observed properties (densities, masses and radii) of the SSCs in NGC\,253 with other SSCs. Red circles represent the SSCs in NGC\,253 from this work. Their marker position is the mean between the upper and lower limit radii and the grey shadow their possible values from the lower to the upper size limit. Black circles are evolved SSCs in NGC\,253 observed by \citet{Watson1996}. Triangles are MW embedded clusters from \citet{Lada2003}. Colored and empty inverted triangles are MW SB and leaky clusters from \citet{Pfalzner2009}, respectively. Squares are SSCs from Local Group galaxies and diamonds are SSCs from outside the Local Group from \citet{Portegies2010}. Panel a) shows the density-radius dependency, with the solid grey line indicating a $r^{-3}$ dependency for SB clusters (all colored symbols) and the dashed grey line a $r^{-4}$ dependency for the leaky clusters (inverted empty triangles). Panel b) shows the mass of these clusters against their radius. 
      }
      \label{fig:rad_dens}
\end{figure}

So far, SSCs have only been observed in external galaxies \citep[see][for a review]{Portegies2010} and they seem to be  the only objects that can become a bound cluster as massive as GCs \citep[][]{Johnson2015}. 
\citet{Pfalzner2009, Pfalzner2011} found that the most massive clusters in the MW evolve in two different sequences: clusters that sustain heavy mass losses expand faster (leaky or unbound clusters) than those that are able to overcome this losses (compact, bound or starburst clusters). Following \citet{Pfalzner2009}, in Fig.~\ref{fig:rad_dens} we have plotted the cluster density ($\text{M}_C/\text{V}_C$) and the cluster mass ($\text{M}_C$), in panels a) and b) respectively, as a function of the cluster radius ($\text{r}_C$)  \citep[][and references therein]{Lada2003, Pfalzner2009, Portegies2010}. The data include NGC\,253 young SSCs from this paper (red circles, where $\text{M}_C=\text{M}_{p*}+\text{M}_{*}$) and \citet{Watson1996} evolved SSCs (black circles), SSCs in galaxies in the Local Group (LG, orange squares) and outside the LG (green diamonds) together. Also plotted are MW leaky (unbound) and starburst (SB, i.e. bound) clusters \citep[inverted triangles][]{Pfalzner2009}.
 For the leaky clusters, the density evolves as $\propto r^{-4}$ (empty inverted triangles and dashed  line in Fig.~\ref{fig:rad_dens}), while the density of SB clusters evolves as  $\propto r^{-3}$ (blue inverted triangles and solid line in Fig.~\ref{fig:rad_dens}). 
This is also illustrated in the  panel b) of Fig.~\ref{fig:rad_dens} where we show how the cluster masses for SB clusters (colored symbols) remain more or less constant with their evolution but leaky clusters (empty symbols) masses changes as they evolve.  The difference seems to be related to a higher SFE in SB clusters than in leaky clusters.
\citet{Pfalzner2013} studied the SFEs required for clusters given a certain density and radius, finding that leaky clusters with SFEs $\sim 20\%$ would not be able to be identified as overdensities after $>5-10$\,Myr  as with this SFE the cluster density declines rapidly. For SB clusters, \citet{Pfalzner2013} find higher SFEs ($\sim 60-70\%$), but higher SFEs ($\geqslant 80\%$) would not explain the observed sizes $>1$\,pc for $>10$\,Myr old clusters. 
Panel a) of Fig.~\ref{fig:rad_dens} shows that young SSCs in NGC\,253 lay in the upper part of the evolutionary sequence of SSCs, i.e. small sizes ($r\lesssim 0.85$\,pc) and high densities
($\sim 10^5-10^6$\,M$_\odot$\,pc$^{-3}$), as expected from young SSCs still unaffected by mechanical feedback. If the SFE is one of the key parameters that determines the survival of a cluster as a bounded system, the high SFEs
($\gtrsim 50\%$) derived for the SSCs detected in NGC\,253 suggests that they could evolve into GCs. But what mechanisms favour such a high SFE  is so far unknown. However, external pressure has to be high enough and it has been proposed to be one of the mechanisms that can maintain high SFRs over enough time to achieve such high SFEs \citep[][and references therein]{Keto2005, Beck2015, Johnson2015}. 

\subsection{On the formation of SSCs}

The physical processes leading to massive star formation from the natal molecular cloud are still not well understood \citep[see][for a review]{Zinnecker2007}.  In fact, the formation and early evolution of the extreme SSCs found in galaxies is one of the most important challenges in the field of star formation.  Several competing theories have been proposed to form  massive stars: i) Monolithic core accretion \citep[][]{McKee2002, McKee2003}; ii) Competitive accretion \citep[proposed by][]{Bonnell2001}.
In the monolithic core accretion, different mechanisms (radiative feedback, gas turbulence and magnetic fields) prevent high fragmentation of the molecular cloud. Then, the densest parts of the cloud have enough material in their surroundings to allow the formation of, at most, a few massive stars. In contrast, in the competitive accretion scenario the cloud fragments and first form a cluster of low-mass stars increasing the cloud gravitational potential well. This helps to accrete the remaining surrounding gas, which is funneled by the low-mass star cluster leading to clustered high-mass stars in the cloud center. Observational evidences supporting the competitive accretion scenario have been found in several massive star-forming regions \citep[e.g.][]{Rivilla2013Orion, Rivilla2013OMC, Rivilla2014}. In a very high density low-mass star cluster, the coalescence of two or more stars might be able to form a more massive star \citep[][]{Bonnell2001}. 
Monolithic core accretion has to face the problem of preventing the further fragmentation of a core in order to be able to form massive stars \citep{Hennebelle2014} making very unlikely the formation of SSCs with high SFEs in very short timescales.  On the other hand, Competitive accretion successfully reproduces the observed stellar Initial Mass Function (IMF) of most MW stellar clusters. Yet, in order to form SSCs like the ones observed in NGC\,253, the phase of the initial low-mass star cluster accretion has to be long enough to accrete enough gas, with large mass accretion rates, to form  a SSC (M$_* \gtrsim 10^5$\,M$_\odot$). Like for the monolithic collapse, the extremely high SFEs and the very short timescales for the SSC formation poses very strong constrains on the timescales for the cluster formation once all the mass has been accreted in a relatively long time scale. This is even more severe for the formation of massive stars by the coalescence of low-mass stars.

The trend found in the SSCs estimated age ($t_\text{age}$), with the more evolved SSCs at smaller projected distances from the galaxy center (see Fig.~\ref{fig:distances}), provides an indication of the recent history of the SSC formation within the inner $160$\,pc of NGC\,253. The  obvious explanation would be that the formation of the SSCs is propagating from the center of the galaxy outwards.

\begin{figure}
\centering
    \includegraphics[width=\linewidth]{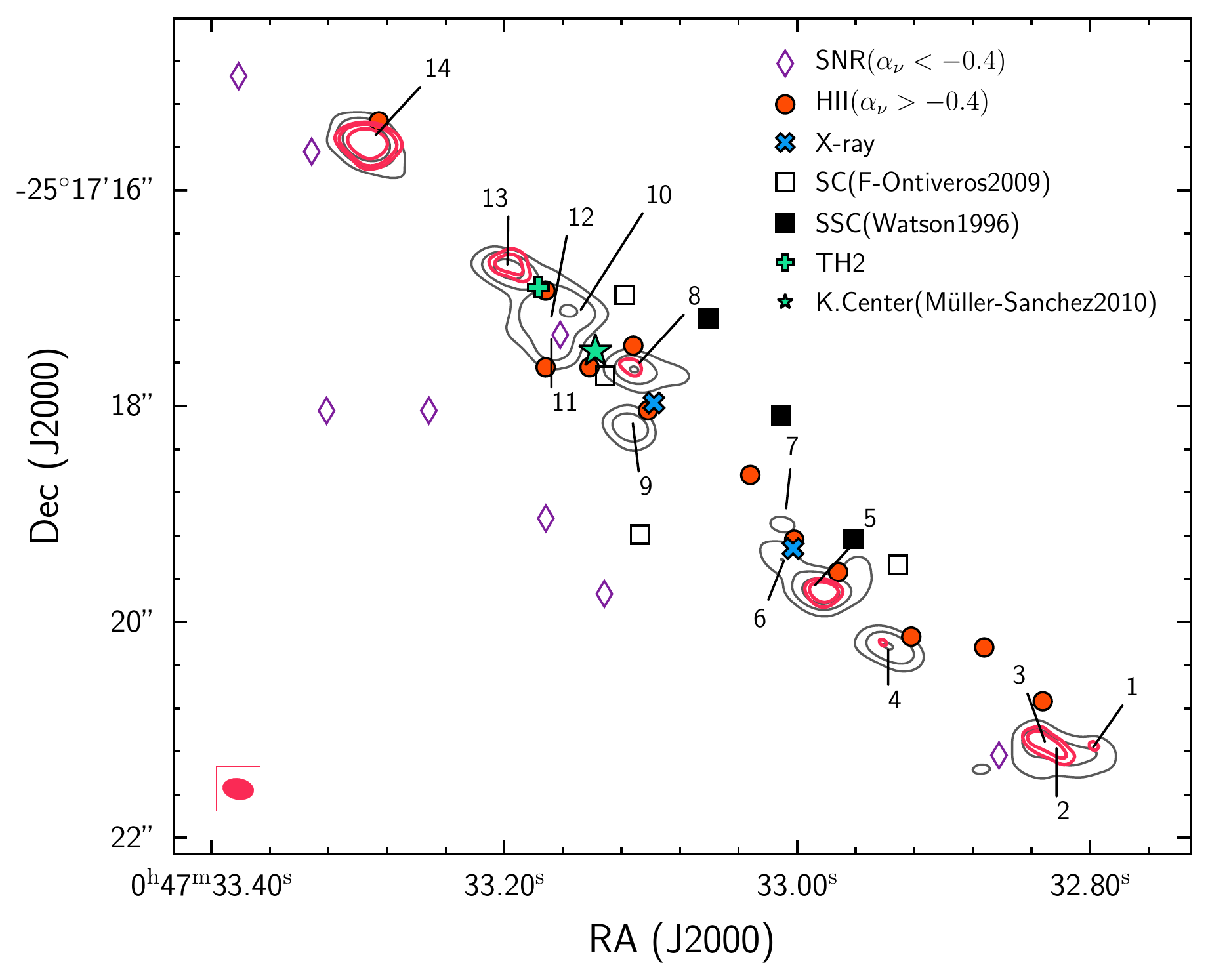}
      \caption{NGC\,253 nuclear region. On grey and red contours are plotted the $218$\,GHz continuum and the HC$_3$N $v_7=1$ $24-23$ emission, respectively. The numbers indicate the position of the forming SSCs discussed in this work. Also indicated are the positions of SNRs with open puple diamonds and \ion{H}{ii} regions with orange circles, studied by \citet[][]{Ulvestad1997}. Open squares show the position of stellar clusters (SCs) observed by \citet[][]{Fernandez-Ont2009} and filled squares the SSCs observed by \citet[][]{Watson1996}. The blue crosses indicate the positions of X-1 and X-2 (coincident with the radio source TH7)  \citep[][]{MullerSanchez2010}. The position of TH2 \citep[][]{Turner1985} and the kinematical center \citep[from][]{MullerSanchez2010} are indicated by a teal plus and star  symbol, respectively. The latter has a $3\sigma\sim 1.2^{\prime \prime}$ uncertainty in its position.
      }
      \label{fig:SSC_pos}
\end{figure}

 Figure~\ref{fig:SSC_pos} shows, together with the young SSCs studied in this work, the location of the Super Nova Remnants (SNR) and \ion{H}{ii} regions observed by \citet{Ulvestad1997} between $1.3-20$\,cm with the VLA; the stellar clusters observed in the IR by \citet{Fernandez-Ont2009}; the more evolved SSCs observed by \citet{Watson1996} with the HST (ages of $5-100 \times 10^6$\,yr); and the positions of two X-ray sources as seen by Chandra \citep{MullerSanchez2010}, along with the position of the  brightest radio source TH2 \citep{Turner1985} and the kinematical center proposed by \citet{MullerSanchez2010}. While most of the \ion{H}{ii} regions are associated to the young SSCs discussed, the SNRs and the old SSCs are located below and above the projected ridge of young SSCs, being plausible that some of the old SSCs and SNRs are located in the spiral arms.

The main properties observed and derived in this work for the SSCs in NGC\,253 (youth, massive, high SFEs and relatively constant SFRs) favours the idea that SSC formation in galaxies represent the most extreme mode of star formation and that it seems to be triggered by external events. Events like galaxy merging, density waves, and mechanical feedback from an active nucleus and/or from star formation will lead to strong shocks which will heat and compress the gas to the sizes and densities required to form the SSCs. In the case of the SSCs observed in NGC\,253, the most likely explanation would be the overpressure produced by hot gas generated by the SN explosion(s) from an early star formation episode in the galaxy center. The trend observed in the age of the SSCs as a function of their location indicates that this might have been produced by a single event. 

\section{Conclusions}

We have used ALMA to study the earliest phases of the formation and evolution of Super Star Clusters (SSCs) which are still deeply embedded in their parental molecular cloud. By using  $0.2^{\prime \prime}$ resolution ($\sim 3$\,pc) ALMA images of the HC$_3$N vibrational excited emission (HC$_3$N$^*$) we have revealed the Super Hot Core (SHC) phase associated with young SSCs (proto-SSCs) in the inner ($160$\,pc) region of the nucleus of the nearby starburst galaxy NGC\,253. Our main results can be summarized as follows:
\begin{description}
    \item[1.] From the $14$ forming SSCs  with strong free-free and dust emission, we have found that $8$ of them show HC$_3$N$^*$ emission (SHC phase), another $4$ show only HC$_3$N emission from the ground state and $2$ of them do not show HC$_3$N emission. 
    \item[2.] We have carried LTE and  non-LTE modelling of the HC$_3$N$^*$ emission to derive the main properties of the SHCs, finding high dust temperatures of  $200-375$\,K and relatively high H$_2$ densities of $1-6 \times 10^6$\,cm$^{-3}$. Somewhat lower temperatures ($\sim 130$\,K) but similar densities are found for the remaining sources with no HC$_3$N$^*$ emission.  We have also estimated, from the lower limit to their sizes, that the LTE and non-LTE  IR luminosities of the SHCs range from  $0.1$ to $1 \times 10^8$\,L$_\odot$.
    \item[3.] The SHCs represent a short lived (a few $10^4$\,yr) phase in the formation of massive stellar clusters, just when protostars are still accreting mass right before massive stars reach the Zero Age Main Sequence (ZAMS) and ionize their surroundings creating Ultra Compact \ion{H}{ii} (UCHII) regions. We have estimated the total stellar mass content of the SSCs in ZAMS stars (M$_*$), from free-free emission inside UCHII regions \citep{Leroy2018}, and in protostars (M$_{p*}$), from the IR luminosities. The derived total masses range from $0.6$ to $10 \times 10^5$\,M$_\odot$. However, the proto/ZAMS luminosity ratio ($\text{L}_{p*}/\text{L}_*$)  in the SSCs shows large variations, of more than two orders of magnitude, from $3.5$ to $<0.01$, indicating that the SSCs are in different evolutionary stages.
   \item[4.] We have then used the $\text{L}_{p*}/\text{L}_*$ ratio  as a clock to measure the evolutionary stage ($t_\text{age}$) of the SSCs. We estimate that the ages of the  youngest SSCs, showing the largest luminosity ratios ($>0.1$), must be a few $10^4$\,yr, and are dominated by the protostar phase (i.e  proto-SSCs). The older ones, with lower luminosity ratios are dominated by the ZAMS phase, are considered ZAMS-SSCs and are likely to be less than $10^6$\,yr since we do not find evidence of mechanical feedback.
   \item[5.] The evolutionary scenario presented above is also supported by the radiative feedback as traced by the HNCO/CS ratio, which measures the degree of photodissociaton of the bulk of the molecular gas in the SSCs. This ratio is systematically higher in the young proto-SSC than in the older ones, as expected if the strong UV radiation from the OB stars in the ZAMS-SSCs has permeated the whole SSC.
    \item[6.] The estimated Star Formation Efficiency (SFE), obtained assuming there has not been significant mass loss (supported by the previous mechanical and radiative feedback analysis) increases from $\sim40\%$ for the proto-SSCs to $>85\%$ for the ZAMS-SSCs. Yet, the gas mass reservoir available for star formation in the proto-SSCs ($1-7 \times 10^5$\,M$_\odot$) is much larger, by nearly one order of magnitude, than in the ZAMS-SSCs ($0.3-7 \times 10^4$\,M$_\odot$), supporting the scenario that star formation is still going on inside the proto-SSCs.
    \item[7.] The SFRs derived for the ZAMS and proto-SSCs phases have similar values, covering a wide range from $0.5$ to $4$\,M$_\odot$\,yr$^{-1}$. For all proto-SSCs we find that the SFR required to achieve a final SFE similar to those of the ZAMS-SSCs ($\sim95\%$)  remains constant during their evolution within a factor of $2$.
    \item[8.] We find a systematic trend  between the estimated age of the SSCs and their projected location in the nuclear region, with the older ZAMS-SSCs located around the center of the galaxy and the  younger proto-SSCs in the outer regions, suggesting an inside-out SSCs formation scenario. We consider that the formation and the high SFE of the SSCs were very likely triggered by the overpressure due to external event(s) that propagates from the inner to the outer nuclear regions.
\end{description}

\section*{Acknowledgements}

We thank the anonymous referee for the suggestions that contributed to improve the paper.
The Spanish Ministry of Science, Innovation and Universities supported this research under grant number ESP2017-86582-C4-1-R, PhD fellowship BES-2016-078808 and MDM-2017-0737 Unidad de Excelencia \quotes{Mar\'ia de Maeztu}.
This paper makes use of the following ALMA data: ADS/JAO.ALMA\#2013.1.00191.S, ADS/JAO.ALMA\#2013.1.00973.S and ADS/JAO.ALMA\#2013.1.00735.S. ALMA is a partnership of ESO (representing its member states), NSF (USA) and NINS (Japan), together with NRC (Canada) and NSC and ASIAA (Taiwan) and KASI (Republic of Korea), in cooperation with the Republic of Chile. The Joint ALMA Observatory is operated by ESO, AUI/NRAO and NAOJ. 
This research made use of Astropy, a community-developed core Python package for Astronomy \citep{Astropy2013}.
V.M.R. has received funding from the European Union's Horizon 2020 research and innovation programme under the Marie Sk\l{}odowska-Curie grant agreement No 664931. 



\bibliographystyle{mnras}
\bibliography{HC3N.bib} 




\appendix




\bsp	
\label{lastpage}
\end{document}